\documentclass[12pt, draftclsnofoot, onecolumn]{IEEEtran}

\input epsf
\usepackage{graphicx}  
\usepackage{graphicx,epstopdf}   
\usepackage{pifont}
\usepackage{bm}
\usepackage{subcaption}
\usepackage{cite}
\usepackage{float}
\usepackage{amsfonts,balance}
\usepackage[utf8x]{inputenc}
\usepackage{cite}
\usepackage{color}

\usepackage[linesnumbered,algoruled,boxed,lined]{algorithm2e} 

\IEEEoverridecommandlockouts
\usepackage{mathtools}
\usepackage{amsthm}
\usepackage{amsmath}
\newtheorem{theorem}{Theorem}

\newtheorem{corollary}{Corollary}

\usepackage{bm,cite,algorithmic,float,amsmath,amssymb}

\usepackage{dsfont}
\usepackage{amssymb}
\usepackage{amsmath}
\usepackage{cite}
\usepackage{citesort}
\usepackage{balance}
\usepackage[utf8x]{inputenc}

\IEEEoverridecommandlockouts

\usepackage{graphicx,epstopdf}
\usepackage{epsfig}	
\usepackage{amsfonts,balance}
\usepackage{bbm}

\usepackage{xcolor,colortbl}
\definecolor{Gray}{gray}{0.9}

\usepackage[
top    = 1.70cm,
bottom = 1.05in,
left   = 0.60 in,
right  = 0.60 in]{geometry}

\begin{document}


\title{Deep Reinforcement Learning for Real-Time Optimization in NB-IoT Networks}

\author{
\IEEEauthorblockN{Nan Jiang, \emph{Student Member, IEEE}, Yansha Deng, \emph{Member, IEEE}, Arumugam Nallanathan, \emph{Fellow, IEEE}, and Jonathon A. Chambers, \emph{Fellow, IEEE}} \\

%
\thanks{}
\thanks{N. Jiang, and A. Nallanathan are with the School of Electronic Engineering and Computer Science, Queen Mary University of London, London E1 4NS, UK (e-mail: \{nan.jiang, a.nallanathan\}@qmul.ac.uk).}
\thanks{Y. Deng is with the Department of Informatics, King's College London, London WC2R 2LS, UK (e-mail: yansha.deng@kcl.ac.uk) (Corresponding author: Yansha Deng).}
\thanks{J. A. Chambers is with the Department of Engineering, University of Leicester, Leicester LE1 7RH, UK (e-mail: jonathon.chambers@le.ac.uk). }

\vspace*{-0.9cm}
}



\maketitle

\vspace*{-1.0cm}

\begin{abstract}
\vspace*{-0.2cm}

NarrowBand-Internet of Things (NB-IoT) is an emerging cellular-based technology that offers a range of flexible configurations for massive IoT radio access from groups of devices with heterogeneous requirements. A configuration specifies the amount of radio resource allocated to each group of devices for random access and for data transmission. Assuming no knowledge of the traffic statistics, there exists an important challenge in ``how to determine the configuration that maximizes the long-term average number of served IoT devices at each Transmission Time Interval (TTI) in an online fashion". Given the complexity of searching for optimal configuration, we first develop real-time configuration selection based on the tabular Q-learning (tabular-Q), the Linear Approximation based Q-learning (LA-Q), and the Deep Neural Network based Q-learning (DQN) in the single-parameter single-group scenario. Our results show that the proposed reinforcement learning based approaches considerably outperform the conventional heuristic approaches based on load estimation (LE-URC) in terms of the number of served IoT devices. This result also indicates that LA-Q and DQN can be good alternatives for tabular-Q to achieve almost the same performance with much less training time. We further advance LA-Q and DQN via Actions Aggregation (AA-LA-Q and AA-DQN) and via Cooperative Multi-Agent learning (CMA-DQN) for the multi-parameter multi-group scenario, thereby solve the problem that Q-learning agents do not converge in high-dimensional configurations. In this scenario, the superiority of the proposed Q-learning approaches over the conventional LE-URC approach significantly improves with the increase of configuration dimensions, and the CMA-DQN approach outperforms the other approaches in both throughput and training efficiency.

\end{abstract}


\vspace*{-0.5cm}
\section{Introduction}
\vspace*{-0.2cm}

To effectively support the emerging massive Internet of Things (mIoT) ecosystem, the 3rd Generation Partnership Project (3GPP) partners have standardized a new radio access technology, namely NarrowBand-IoT (NB-IoT) \cite{RohdeSchwarz2016white}. NB-IoT is expected to provide reliable wireless access for IoT devices with various types of data traffic, and to meet the requirement of extended coverage. As most mIoT applications favor delay-tolerant data traffic with small size, such as data from alarms, and meters, monitors, the key target of NB-IoT design is to deal with the sporadic uplink transmissions of massive IoT devices \cite{dhillon2017wide}.

NB-IoT is built from legacy Long-Term Evolution (LTE) design, but only deploys in a narrow bandwidth (180 KHz) for Coverage Enhancement (CE) \cite{wang2017primer}. Different from the legacy LTE, NB-IoT only defines two uplink physical channel resource to perform all the uplink transmission, including the Random Access CHannel (RACH) resource (i.e., using NarrowBand Physical Random Access CHannel, a.k.a. NPRACH) for RACH preamble transmission, and the data resource (i.e., using NarrowBand Physical Uplink Shared CHannel, a.k.a. NPUSCH) for control information and data transmission. To support various traffic with different coverage requirements, NB-IoT supports up to three CE groups of IoT devices sharing the uplink resource in the same band. Each group serves IoT devices with different coverage requirements distinguishing based on a same broadcast signal from the evolved Node B (eNB) \cite{wang2017primer}. At the beginning of each uplink Transmission Time Interval (TTI), eNB selects a system configuration that specifies the radio resource allocated to each group in order to accommodate the RACH procedure along with the remaining resource for data transmission. The key challenge is to optimally balance the allocations of channel resource between the RACH procedure and data transmission so as to provide maximum success accesses and transmissions in massive IoT networks. Allocating too many resource for RACH enhances the random access pernformace, while leaving insufficient resource for data transmission.

Unfortunately, dynamic RACH and data transmission resource configuration optimization is an untreated problem in NB-IoT. Generally speaking, the eNB observes the transmission receptions of both RACH (e.g., number of successfully received preambles and collisions) and data transmission (e.g., number of successful scheduling and unscheduling) for all groups at the end of each TTI. This historical information can be potentially used to predict traffic from all groups and to facilitate the optimization of future TTIs' configurations. Even if one knew all the relevant statistics, tackling this problem in an exact manner would result in a Partially Observable Markov Decision Process (POMDP) with large state and action spaces, which would be generally intractable. The complexity of the problem is compounded by the lack of a prior knowledge at the eNB regarding the stochastic traffic and unobservable channel statistics (i.e., random collision, and effects of physical radio including path-loss and fading). The related works will be briefly introduced in the following two subsections.

\subsubsection{Related works on real-time optimization in cellular-based networks}

In light of this POMDP challenge, prior works \cite{wiriaatmadja2015hybrid,duan2016d} studied real-time resource configuration of RACH procedure and/or data transmission by proposing dynamic Access Class Barring (ACB) schemes to optimize the number of served IoT devices. These optimization problems have been tackled under the simplified assumptions that at most two configurations are allowed and that the optimization is executed for a single group without considering errors due to wireless transmission. In order to consider more complex and practical formulations, Reinforcement Learning (RL) emerges as a natural solution given its capability in interacting with the practical environment and feedback in the form of the number of successful and unsuccessful transmissions per TTI. Distributed RL based on tabular Q-learning (tabular-Q) has been proposed in \cite{bello2014application,chu2012aloha,yan2013distributed,naddafzadeh2010distributed}. In \cite{bello2014application,chu2012aloha,yan2013distributed}, the authors studied distributed tabular-Q in slotted-Aloha networks, where each device learns how to avoid collisions by finding a proper time slot to transmit packets. In \cite{naddafzadeh2010distributed}, the authors implemented tabular-Q agents at the relay nodes for cooperatively selecting its transmit power and transmission probability to optimize the total number of useful received packets per consumed energy. Centralized RL has also been studied in \cite{chen2004q,ihun2017A,Luis2018Reinforcement}, where the RL agent was implemented at the base station site. In \cite{chen2004q}, a learning-based scheme was proposed for radio resource management in multimedia wide-band code-division multiple access systems to improve spectrum utilization. In \cite{ihun2017A,Luis2018Reinforcement}, the authors studied the tabular-Q based ACB schemes in cellular networks, where a Q-agent was implemented at an eNB aiming at selecting the optimal ACB factor to maximize the access success probability of RACH procedure.

\subsubsection{Related works on optimization in NB-IoT}
In NB-IoT networks, most existing studies either focused on the resource allocation during RACH procedure \cite{harwahyu2018optimization,oh2017efficient}, or that during the data transmission \cite{malik2018radio,yu2017uplink}. For RACH procedure, the access success probability was statistically optimized in \cite{harwahyu2018optimization} using exhaustive search, and the authors in \cite{oh2017efficient} studied the fixed-size data resource scheduling for various resource requirements. For the data transmission, \cite{malik2018radio} presented an uplink data transmission time slot and power allocation scheme to optimize the overall channel gain, and \cite{yu2017uplink} proposed a link adaptation scheme, which dynamically selects modulation and coding level, and the repetition value according to the acknowledgment/negative-acknowledgment feedback to reduce the uplink data transmission block error ratio. More importantly, these works ignore the time-varied heterogeneous traffic of massive IoT devices, and considered a snap shot \cite{harwahyu2018optimization,malik2018radio,yu2017uplink} or steady-state behavior \cite{oh2017efficient} of NB-IoT networks. Our most relevant work is \cite{azari2018latency}, where the authors studied the steady-state behavior of NB-IoT networks from the perspective of a single device. Optimizing some of the parameters of the NB-IoT configuration, namely the repetition value (to be defined below) and time intervals between two consecutive scheduling of NPRACH and NPDCCH, was carried out in terms of latency and power consumption in \cite{azari2018latency} using a queuing framework.

Unfortunately, the tabular-Q framework in \cite{ihun2017A,Luis2018Reinforcement} cannot be used to solve the multi-parameter multi-group optimization problem in uplink resource configuration of NB-IoT networks, due to their incapability to address high-dimensional state space and variable selection. More importantly, whether their proposed RL-based resource configuration approaches \cite{ihun2017A,Luis2018Reinforcement} outperform the conventional resource configuration approaches \cite{duan2016d,wiriaatmadja2015hybrid} is still unknown. In this paper, we develop RL-based uplink resource configuration approaches to dynamically optimize the number of served IoT devices in NB-IoT networks. To showcase the efficiency, we compare the proposed RL-based approaches with the conventional heuristic uplink resource allocation approaches. The contributions can be summarized as follows:
\begin{itemize}
\item We develop an RL-based framework to optimize the number of served IoT devices by adaptively configuring uplink resource in NB-IoT networks. The uplink communication procedure in NB-IoT is simulated by taking into account the heterogeneous IoT traffics, the CE group selection, the RACH procedure, and the uplink data transmission resource scheduling. This generated simulation environment is used for training the RL-based agents before deployment, and these agents will be updated according to the real traffic in practical NB-IoT networks in an online manner.
\item
We first study a simplified NB-IoT scenario considering the single parameter and the single CE group, where a basic tabular-Q was developed to compare with the revised conventional Load Estimation based Uplink Resource Configuration (LE-URC) scheme. The tabular-Q is further advanced by implementing function approximators with different computational complexities, namely, Linear Approximator (LA-Q) and Deep Neural Networks (Deep Q-Network, a.k.a. DQN) to elaborate their capability and efficiency in dealing with high-dimensional state space. 
\item 
We then study a more practical NB-IoT scenario with multiple parameters and multiple CE groups, where direct implementation of the LA-Q or DQN is not feasible due to the increasing size of the parameter combinations. To solve it, we propose Action Aggregation approaches based on LA-Q and DQN, namely, AA-LA-Q and AA-DQN, which guarantee convergence capability by sacrificing certain accuracy in the parameters selection. Finally, a Cooperative Multi-Agent learning based on DQN (CMA-DQN) is developed to break down the selection in high-dimensional parameters into multiple parallel sub-tasks by using that a number of DQN agents are cooperatively trained to produce each parameter for each CE group.
\item 
In the simplified scenario, our results show that the number of served IoT devices with tabular-Q considerably outperforms that with LE-URC, while LA-Q and DQN achieve almost the same performance as that of tabular-Q using much less training time. In the practical scenario, the superiority of Q-learning based approaches over LE-URC significantly improves. Especially, CMA-DQN outperforms all other approaches in terms of both throughput and training efficiency, which is mainly due to the use of DQN enabling operation over a large state space and the use of multiple agents dealing with the large dimensionality of parameters selection. 
\end{itemize}

The rest of the paper is organized as follows. Section II provides the problem formulation and system model. Section III illustrates the preliminary and the conventional LE-URC. Section IV proposes Q-leaning based uplink resource configuration approaches in the single-parameter single-group scenario. Section V presents the advanced Q-learning based approaches in the multi-parameter multi-group scenario. Section VI elaborates the numerical results, and finally, Section VII concludes the paper.

\vspace*{-0.3cm}
\section{Problem Formulation and System Model}
\vspace*{-0.1cm}

\begin{figure}[htbp!]
\setlength{\abovecaptionskip}{0pt}
    \begin{center}
        \includegraphics[width=0.9\textwidth]{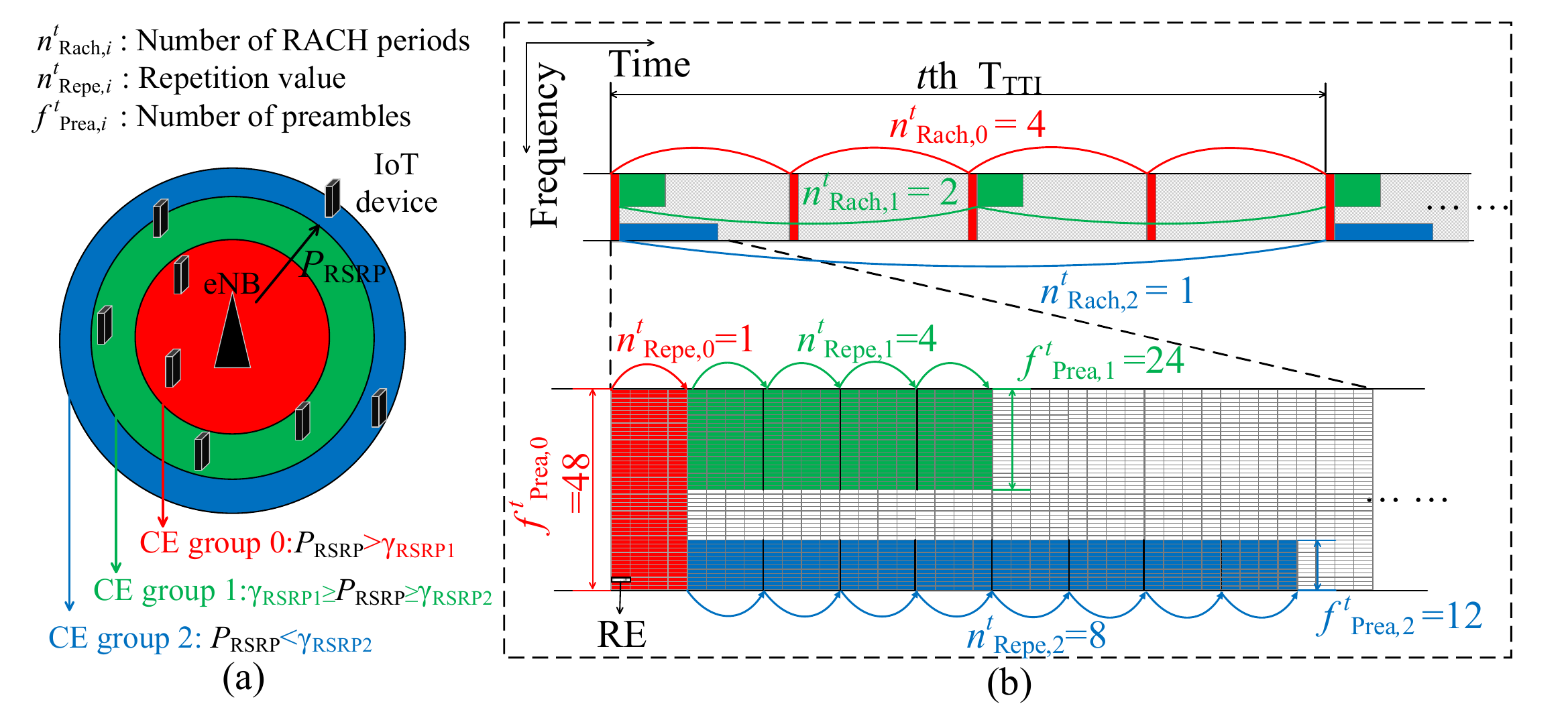}
        \caption{(a) Illustration of system model; (b) Uplink channel frame structure.
        }\label{fig:structure}
    \end{center}
    \vspace*{-0.6cm}
\end{figure}

As illustrated in Fig. \ref{fig:structure}(a), we consider a single-cell NB-IoT network composed of an eNB located at the center of the cell, and a set of static IoT devices randomly located in an area of the plane $\mathbb R^2$, and remain spatially static once deployed. The devices are divided into three CE groups as further discussed below, and the eNB is unaware of the status of these IoT devices, hence no uplink channel resource is scheduled to them in advance. In each IoT device, uplink data is generated according to random inter-arrival processes over the TTIs, which are Markovian and possibly time-varying.

\vspace*{-0.1cm}
\captionsetup{singlelinecheck=false}  

\subsection{Problem Formulation}
\vspace*{-0.25cm}
With packets waiting for service, an IoT device executes the contention-based RACH procedure in order to establish a Radio Resource Control (RRC) connection with the eNB. The contention-based RACH procedure consists of four steps, where an IoT device transmits a randomly selected preamble, for a given number of times according to the repetition value $n^t_{\text{Repe},i}$ \cite{RohdeSchwarz2016white}, to initial RACH procedure in step 1, and exchanges control information with the eNB in the next three steps \cite{LTE2013dahlman}. The RACH process can fail if: (\textit{i}) a collision occurs when two or more IoT devices selecting the same preamble; or (\textit{ii}) there is no collision, but the eNB cannot detect a preamble due to low SNR. Note that a collision can be still detected in step 3 of RACH when the collided preambles are not detected in step 1 of RACH following 3GPP report \cite{3GPP2011Study}. This assumption is different from our previous works \cite{jiang2018rach,jiang2018collision}, which only focuses on the preamble detection analysis in step 1 of RACH.

As shown in Fig. 1(b), for each TTI $t$ and for each CE group $i=0,1,2$, in order to reduce the chance of a collision, the eNB can increase the number $n^t_{\text{Rach},i}$ of RACH periods in the TTI or the number $f^t_{\text{Prea},i}$ of preambles available in each RACH period \cite{3GPP2017PhyCM}. Furthermore, in order to mitigate the SNR outage, the eNB can increase the number $n^t_{\text{Repe},i}$ of times that a preamble transmission is repeated by a device in group $i$ in one RACH period \cite{3GPP2017PhyCM} of the TTI.

After the RRC connection is established, the IoT device requests uplink channel resource from the eNB for control information and data transmission. As shown in Fig. 1(b), given a total number of resource $R_\text{Uplink}$ for uplink transmission in the TTI, the number of available resource for data transmission $R^t_\text{DATA}$ is written as $R^t_\text{DATA} = R_\text{Uplink}- R^t_\text{RACH}$, where $R^t_\text{RACH}$ is the overall number of Resource Elements (REs)\footnote{The uplink channel consists of 48 sub-carriers within 180 kHz bandwidth. With a 3.75 kHz tone spacing, one RE is composed of one time slot of 2 ms and one sub-carrier of 3.75 kHz \cite{RohdeSchwarz2016white}. Note that the NB-IoT also supports 12 sub-carriers with 15 kHz tone spacing for NPUSCH, but NPRACH only supports 3.75 kHz tone spacing \cite{RohdeSchwarz2016white}.} allocated for the RACH procedure. This can be computed as $R^t_\text{RACH} = B_\text{RACH} \sum_{i=0}^{2} n_{\text{Rach},i} n_{\text{Repe},i}  f_{\text{Prea},i}$, where $B_\text{RACH}$ measures the number of REs required for one preamble transmission.

In this work, we tackle the problem of optimizing the RACH configuration defined by parameters $A^t=\{n^t_{\text{Rach},i},f^t_{\text{Prea},i},$ $n^t_{\text{Repe},i}\}_{i=0}^{2}$ for each $i$th group in an online manner for every TTI $t$. In order to make this decision at the beginning of every TTI $t$, the eNB accesses all prior history $U^{t'}$ in TTIs $t'=1,...,t-1$ consisting of the following variables: the number of the collided preambles $V^{t'}_{{\rm cp},i}$, the number of the successfully received preambles $V^{t'}_{{\rm sp},i}$, and the number of idle preambles $V^{t'}_{{\rm ip},i}$ of the $i$th CE group in the $t$th TTI for the RACH, as well as the number of IoT devices that have successfully sent data $V^{t'}_{{\rm su},i}$ and the number of IoT devices that are waiting for being allocated data resource $V^{t'}_{{\rm un},i}$. We denote $O^{t}=\{A^{t-1}, U^{t-1}, A^{t-2}, U^{t-2},\cdots,A^{1}, U^{1}\}$ as the observed history of all such measurements and past actions. 

The eNB aims at maximizing the long-term average number of devices that successfully transmit data with respect to the stochastic policy $\pi$ that maps the current observation history $O^{t}$ to the probabilities of selecting each possible configuration $A^t$. This problem can be formulated as the optimization
\vspace*{-0.1cm}
\begin{align}\label{q6-1}
&(\text{P1}):  \mathop {\textbf{max}}\limits_{ \{\pi(A^t|O^t)\}}     \quad   \sum_{k=t}^{\infty} \sum_{i=0}^2  \gamma^{k-t} {\mathbb E}_{\pi} [ V^{k}_{{\rm su},i} ],
\end{align}
where $\gamma \in [0,1)$ is the discount rate for the performance in future TTIs and index $i$ runs over the CE groups. Since the dynamics of the system is Markovian over the TTI and is defined by the NB-IoT protocol to be further discussed below, this is a POMDP problem that is generally intractable. Approximate solutions will be discussed in Sections III, IV, and V.

\subsection{NB-IoT Access Network}

\begin{figure}[htbp!]
\setlength{\abovecaptionskip}{0pt}
\vspace*{-0.3cm}
    \begin{center}
        \includegraphics[width=1\textwidth]{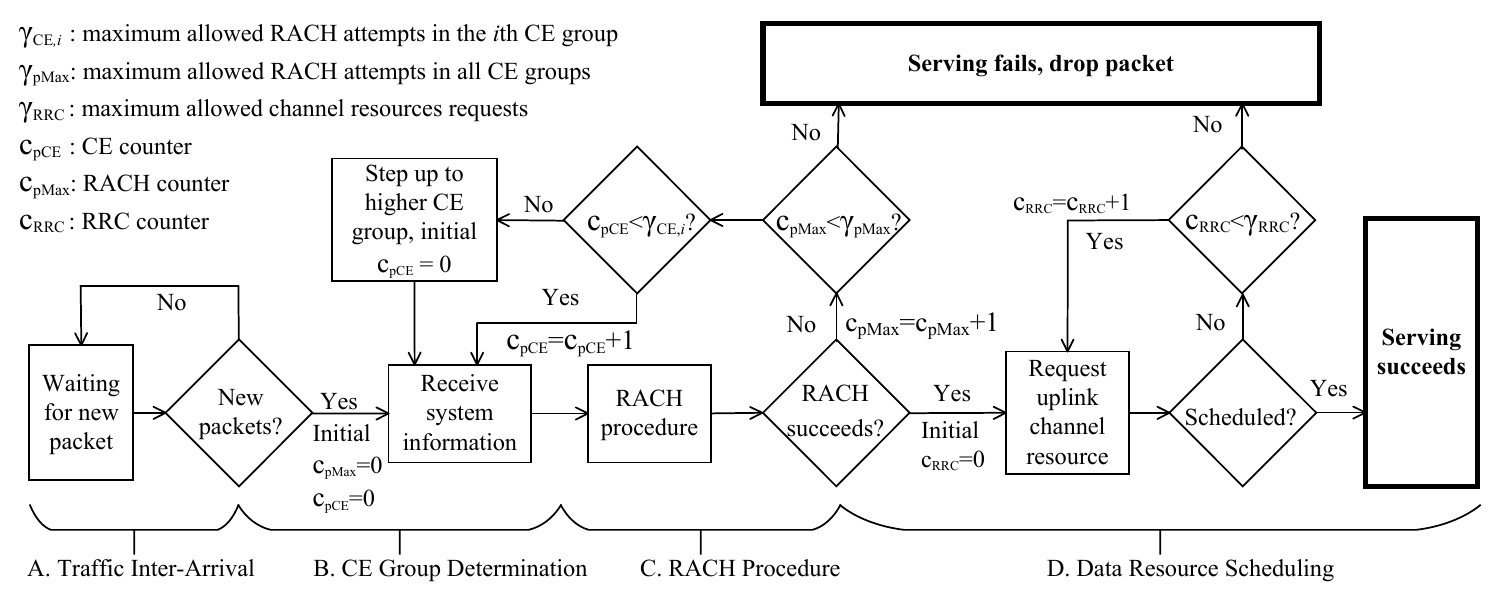}
        \caption{Uplink data transmission procedure from the perspective of an IoT device in NB-IoT networks.
        }\label{fig:CProcedure}
    \end{center}
    \vspace*{-0.9cm}
\end{figure}

We now provide additional details on the model and on the NB-IoT protocol. To capture the effects of the physical radio, we consider the standard power-law path-loss model that the path-loss attenuation is ${ u }^{ - \eta  }$, with the propagation distance ${ u }$ and the path-loss exponent ${\eta}$. The system is operated in a Rayleigh flat-fading environment, where the channel power gains $h$ are exponentially distributed (i.i.d.) random variables with unit mean. Fig. \ref{fig:CProcedure} presents the uplink data transmission procedure from the perspective of an IoT device in NB-IoT networks, which consists of the four stages that are explained in the following four subsections to introduce the system model.

\captionsetup{singlelinecheck=false}  

\vspace*{-0.0cm}
\subsubsection{Traffic Inter-Arrival}
\vspace*{-0.0cm}

We consider two types of IoT devices with different traffic models, including periodical traffic and bursty traffic, which is a heterogeneous traffic scenario for diverse IoT applications \cite{shafiq2012first,kim2014m2m}. The periodical traffic coming from periodic uplink reporting tasks, such as metering or environmental monitoring, is the most common traffic model in NB-IoT networks \cite{3GPP2015Cellular}. The bursty traffic due to emergency events, such as fire alarms and earthquake alarms, captures the complementary scenario in which a massive number of IoT devices tries to establish RRC connection with the eNB \cite{3GPP2011Study}. Due to the nature of slotted-Aloha, an IoT device can only transmit a preamble at the beginning of a RACH period, which means that IoT devices executing RACH in a RACH period comes from those who received an inter-arrival within the interval between with the last RACH period. For the periodical traffic, the first packet is generated using Uniform distribution over $T_\text{periodic}$ (ms), and then repeated every $T_\text{periodic}$ ms. The packet inter-arrival rate measured in each RACH period at each IoT device is hence expressed by
\vspace*{-0.0cm}
\begin{align}\label{q7}
\mu^t_\text{period} =  \frac{T_\text{TTI}}{n^t_{\text{Rach},i}} \times \frac{1} {T_\text{periodic}},
\end{align}
where $n^t_{\text{Rach},i}$ is the number of RACH periods in the $t$th TTI, $\frac{T_\text{TTI}}{n^t_{\text{Rach},i}}$ is the duration between neighboring RACH periods. The bursty traffic is generated within a short period of time $T_\text{bursty}$ starting from a random time $\tau_0$. The traffic instantaneous rate in packets in a period is described by a function $p(\tau)$ so that the packets arrival rate in the $j$th RACH period of the $t$th TTI is given by
\vspace*{-0.1cm}
\begin{align}\label{q8}
\mu_\text{bursty}^{t,j} = \int_{{{\rm{\tau }}_{j-1}}}^{{\tau_{j}}} {p(\tau)} d\tau ,
\end{align}
where $\tau_j$ is the starting time of the $j$th RACH period in the $t$th TTI, $\tau_j - \tau_{j-1} = \frac{T_\text{TTI}}{n^t_{\text{Rach},i}}$, and the distribution $p(\tau)$ follows the time limited Beta profile given as \cite[Section 6.1.1]{3GPP2011Study}
\vspace*{-0.1cm}
\begin{align}\label{q9}
p(\tau) = \frac{{{\tau^{{\alpha}-1}}{{(T_\text{bursty} - \tau)}^{{\beta}-1}}}}{{{{T_\text{bursty}}^{{\alpha} + {\beta} - 2}}\text{Beta}({\alpha} , {\beta} )}} ,
\end{align}
In (\ref{q9}), $\text{Beta}({\alpha} , {\beta})$ is the Beta function with the constant parameters ${\alpha}$ and ${\beta}$ \cite{gupta2004handbook}.

\vspace*{-0.0cm}
\subsubsection{CE Group Determination}
\vspace*{-0.0cm}

 Once an IoT device is backlogged, it first determines its associated CE group by comparing the received power of the broadcast signal $P_\text{RSRP}$ to the Reference Signal Received Power (RSRP) thresholds $\{\gamma_\text{RSRP1}, \gamma_\text{RSRP2}\}$ according to the rule \cite{3GPP2017Physical}
 \vspace*{-0.1cm}
\begin{align}\label{q3}
 \left\{ \begin{aligned} 
   & \text{CE group 0,} & &\text{if } P_\text{RSRP} > \gamma_\text{RSRP1} ,
 \\& \text{CE group 1,}  & &\text{if } \gamma_\text{RSRP1} \ge P_\text{RSRP} \ge \gamma_\text{RSRP2} ,
 \\& \text{CE group 2,}  & &\text{if } P_\text{RSRP} < \gamma_\text{RSRP2} .
\end{aligned}  \right. 
\end{align}

In (\ref{q3}), the received power of broadcast signal $P_\text{RSRP}$ is expressed as
\vspace*{-0.3cm}
\begin{align}\label{q2}
P_\text{RSRP} = P_\text{NPBCH}  u  ^{ -\eta  },
\end{align}
where $u$ is the device's distance from the eNB, and $P_\text{NPBCH}$ is the broadcast power of eNB  \cite{3GPP2017Physical}. Note that $P_\text{RSRP}$ is obtained by averaging the small-scale Rayleigh fading of the received power \cite{3GPP2017Physical}.

\vspace*{-0.0cm}
\subsubsection{RACH Procedure}
\vspace*{-0.0cm}
After CE group determination, each backlogged IoT device in group $i$ repeats a randomly selected preamble $n^t_{\text{Repe},i}$ times in the next RACH period by using a pseudo-random frequency hopping schedule. The pseudo-random hopping rule is based on the current repetition time as well as the Narrowband Physical Cell ID, and in one repetition, a preamble consists of four symbol groups, which are transmitted with fixed size frequency hopping \cite{lin2016random,jiang2018rach,RohdeSchwarz2016white}. Therefore, a preamble is successfully detected if \textit{at least} one preamble repetition succeeds, which in turn happens if \textit{all} of its four symbol groups are correctly decoded \cite{jiang2018rach}. Assuming that correct detecting is determined by the SNR level $\text{SNR}^t_{\text{sg},j,k}$ for the $j$th repetition and the $k$ symbol group, the correct detecting event $S_{\text{pd}}$ can be expressed as
\vspace*{-0.2cm}
\begin{align}\label{q06-4}
S_{\text{pd}}  \buildrel \Delta \over =  \bigcup\limits_{j=1}^{n^t_{\text{Repe},i}} { \Big(  \bigcap\limits_{k=1}^{4}  \big\{ {\text{SNR}^t_{\text{sg},j,k}} \ge \gamma_{\text th} \big\} \Big) }   ,
\end{align}
where $k$ is the index of symbol group in the $j$th repetition, ${n^t_{\text{Repe},i}}$ is the repetition value of the $i$th CE group in the $t$th TTI, $\big\{ {\text{SNR}^t_{\text{sg},j,k}} \ge \gamma_{\text th} \big\}$ means that the preamble symbol group is successfully decoded when its received SNR ${\text{SNR}^t_{\text{sg},j,k}}$ above a threshold $\gamma_{\text th}$, and ${\text{SNR}^t_{\text{sg},j,k}}$ is expressed as
\vspace*{-0.3cm}
\begin{align}\label{q06}
{\text{SNR}^t_{\text{sg},j,k}} = { P_{\text{RACH},i} {u}^{-\eta} {h}}  /  {\sigma ^2}.
\end{align}
In (\ref{q06}), ${u}$ is the Euclidean distance between the IoT device and the eNB, $\eta $ is the path-loss attenuation factor, $h$ is the Rayleigh fading channel power gain from the IoT device to the eNB, ${\sigma ^2}$ is the noise power, and $P_{\text{RACH,}i}$ is the preamble transmit power in the $i$th CE group defined as
\vspace*{-0.2cm}
\begin{align}\label{q1}
P_{\text{RACH,}i} =  \left\{ \begin{aligned}  
  & \text{min } \{ {\rho{ u }^{ \eta  }} , \text{ } P_\text{RACHmax} \} ,  & & \hspace*{-0.1cm} i = 0, 
 \\
  & P_\text{RACHmax},  &  & \hspace*{-0.1cm} i = 1 \text{ or } 2 .
\end{aligned} \right. 
\end{align}
where $i$ is the index of CE groups, IoT devices in the CE group 0 ($i = 0$) transmit preamble using the full path-loss inversion power control \cite{3GPP2017Physical}, which maintains the received signal power at the eNB from IoT devices with different distance equalling to the same threshold $\rho$, and $P_\text{RACHmax}$ is the maximal transmit power of an IoT device. The IoT devices in the CE group 1 and group 2 always transmit preamble using the maximum transmit power \cite{3GPP2017Physical}.

As shown in the RACH procedure of Fig. \ref{fig:CProcedure}, if a RACH fails, the IoT device reattempts the procedure until receiving a positive acknowledgement that RRC connection is established, or exceeding $\gamma_{\text{pCE},i}$ RACH attempts while being part of one CE group. If these attempts exceeds $\gamma_{\text{pCE},i}$, the device switches to a higher CE group if possible \cite{3GPP2017MAC}. Moreover, the IoT device is allowed to attempt the RACH procedure no more than $\gamma_{\text{pMax}}$ times before dropping its packets. These two features are counted by $c_{\text{pCE}}$ and $c_{\text{pMax}}$, respectively.

\vspace*{-0.0cm}
\subsubsection{Data Resource Scheduling}
\vspace*{-0.0cm}

After the RACH procedure succeeds, the RRC connection is successfully established, and the eNB schedules resource from the data channel resource $R^t_\text{DATA}$ to the associated IoT device for control information and data transmission as shown in Fig \ref{fig:structure}(b).
To allocate data resource among these devices, we adopt a basic random scheduling strategy, whereby an ordered list of all devices that have successfully completed the RACH procedure but have not received a data channel is compiled using a random order. In each TTI, devices in the list are considered in order for access to the data channel until the data resource are insufficient to serve the next device in the list. The remaining RRC connections between the unscheduled IoT devices and the eNB will be preserved within at most $\gamma_\text{RRC}$ subsequent TTIs counting by $c_\text{RRC}$, and attempts will be made to schedule the device's data during these TTIs \cite{3GPP2017Requirements,3GPP2017MAC}. The condition that the data resource are sufficient in TTI $t$ is expressed as
\vspace*{-0.1cm}
\begin{align}\label{q5-1}
{R}^t_{\text{DATA}} \ge \sum_{i=0}^{2} {r}^t_{\text{DATA},i} V^t_{\text{sch},i},
\end{align}
where $\sum_{i=0}^{2} V^t_{\text{sch},i}\leq \sum_{i=0}^{2} (V^t_{\text{sp},i} + V^{t-1}_{\text{un},i})$ is the number of scheduled devices limited by the upper bound denoted by IoT devices with successful RACH $V^t_{\text{sp},i}$ in the current TTI $t$ as well as unscheduled IoT devices $V^{t-1}_{\text{un},i}$ in the last TTI $(t-1)$, ${r}^t_{\text{DATA},i} = B_\text{DATA}  \times  n^t_{\text{Repe},i}$ is the number of required REs for serving one IoT device within the $i$th CE group, and $B_\text{DATA}$ is the number of REs per repetition for control signal and data transmission\footnote{The basic scheduling unit of NPUSCH is resource unit (RU), which has two formats. NPUSCH format 1 (NPUSCH-1) is with 16 REs for data transmission, and NPUSCH format 2 (NPUSCH-2) is with 4 REs for carrying control information \cite{wang2017primer,3GPP2017PhyCM}.}. Note that $n^t_{\text{Repe},i} $ is the repetition value for the $i$th CE group in the $t$th TTI, which is the same as for preamble transmission \cite{RohdeSchwarz2016white}.

\vspace*{-0.2cm}
\section{Preliminary and Conventional Solutions}
\subsection{Preliminary}

The optimized number of served IoT devices over the long term given in Eq. (\ref{q6-1}) is really complicated, which cannot be easily solved via the conventional uplink resource approach. Therefore, most prior works simplified the objective to dynamically optimize the single parameter to achieve the maximum number of served IoT devices in the single group without consideration of future performance \cite{duan2016d,wiriaatmadja2015hybrid}, which is expressed as
\begin{align}\label{q6-2}
&(\text{P2}):  \quad \mathop {\text{max}}\limits_{ \pi (x \big| O^t)}   \quad  {\mathbb E}_{\pi} [ V^{t}_{{\rm su},0} ],
\end{align}
where $x$ is the optimized single parameter.

To maximize number of served IoT devices in the $t$th TTI, the configuration $x$ is expected to be dynamically adjusted according to the actual number of IoT devices that will execute RACH attempts $D^t_{\text{RACH}}$, which refers to the current load of the network. Note that in practice, this load information is unable to be detected at the eNB. Thus, it is necessary to estimate the load based on the previous transmission reception from the $1$th to $(t-1)$th TTI $O^t$ before the uplink resource configuration in the $t$th TTI.

In \cite{duan2016d}, the authors designed a dynamic ACB scheme to optimize the problem given in Eq. (\ref{q6-1}) via adjusting the ACB factor.
The ACB factor is adapted based on the knowledge of traffic load, which is estimated via moment matching. The estimated number of RACH attempting IoT devices in the $t$th TTI ${\hat D}^t_{\text{RACH}}$ is expressed as: 
\begin{align}\label{q6-3}
&{\hat D}^t_{\text{RACH}} = \text{max } \Big\{    0,    {\hat D}^{t-1}_{\text{RACH}}  + \text{max }   \big\{ -f^{t-1}_{\text{Prea},0}, {\hat D}^{t}_{\text{RACH}} - {\hat D}^{t-1}_{\text{RACH}}  \big\}      \Big\}
\end{align}
where $f^{t-1}_{\text{Prea},0}$ is the number of allocated preambles in the $(t-1)$th TTI, and ${\hat D}^{t-1}_{\text{RACH}}$ is the estimated number of devices performing RACH attempts in the $(t-1)$th TTI given as
\begin{align}\label{q6-33}
{\hat D}^{t-1}_{\text{RACH}} =  f^{t-1}_{\text{Prea},0} / \Big[  \text{min} \Big\{ 1, p_\text{ACB}^{t-1} \big(  1+\frac{(V_{\text{cp},0}^{t-1}-u_{M,p^*})e}{2f^{t-1}_{\text{Prea},0}})  \big)^{-1} \Big\} \Big] . 
\end{align}
In Eq. (\ref{q6-33}), $p_\text{ACB}^{t-1}$, $f^{t-1}_{\text{Prea},0}$, and $V_{\text{cp},0}^{t-1}$ are the ACB factor, the number of preambles and the observed number of collided preambles in the $(t-1)$th TTI, and $u_{M,p^*}$ is an estimated factor given in \cite[Eq. (32)]{duan2016d}.

In Eq. (\ref{q6-3}), ${\hat D}^{t}_{\text{RACH}} - {\hat D}^{t-1}_{\text{RACH}} \approx {\hat D}^{t-1}_{\text{RACH}} - {\hat D}^{t-2}_{\text{RACH}}$ is the difference between the estimated numbers of RACH requesting IoT devices in the $(t-1)$th and the $t$th TTIs, which is obtained by assuming that the number of successful RACH IoT devices does not change significantly in these two TTIs \cite{duan2016d}.

This dynamic control approach is designed for an ACB scheme, which is only triggered when the exact traffic load is bigger than the number of preambles (i.e., $D^t_{\text{RACH}} > f^t_{\text{Prea},0}$). Accordingly, the related backlog estimation approach is only used when $D^t_{\text{RACH}} > f^t_{\text{Prea},0}$. However, it cannot estimate the load when $D^t_{\text{RACH}} < f^t_{\text{Prea},0}$, which is required in our problem.

\subsection{Resource Configuration in Single Parameter Single CE Group Scenario}
In this subsection, we modify the load estimation approach given in \cite{duan2016d} via estimating based on the last number of the collided preambles $V^{t-1}_{\text{cp},0}$ and the previous numbers of idle preambles $V^{t-1}_{\text{ip},0}, V^{t-2}_{\text{ip},0},\cdots $. And then, we propose an uplink resource configuration approach based on this revised load estimation, namely, LE-URC.

\subsubsection{Load Estimation}

By definition, ${\boldsymbol{\cal F}}_\text{Prea}$ is the set of valid number of preambles that the eNB can choose, where each IoT device selects a RACH preamble from $f^t_{\text{Prea},0}$ available preambles with an equal probability given by $1 / f^t_{\text{Prea},0}$. For a given preamble $j$ transmitted to the eNB, let $d_j$ denotes the number of IoT devices that selects the preamble $j$. The probability that no IoT device selects preamble $j$ is
\begin{align}\label{aa1}
{\mathbb P} \{ d_j = 0 \big| D^{t-1}_{\text{RACH},0} = n  \}=   \big(1-\frac{1}{f^{t-1}_{\text{Prea},0}} \big)^{n}.
\end{align}
The expected number of preambles experiencing idles ${\mathbb E} \{ {\cal V}^{t-1}_{{\rm idle},0} \big| D^{t-1}_{\text{RACH},0}  = n  \}$ in the $(t-1)$th TTI is given by
\vspace*{-0.63cm}
\begin{align}\label{aa2}
& {\mathbb E} \{ {\cal V}^{t-1}_{{\rm ip},0} \big| D^{t-1}_{\text{RACH},0}  = n  \}  = \Big( \sum\limits_{j>1}^{f^{t-1}_{\text{Prea},0}} {\mathbb P} \{ d_j = 0 \big| D^{t-1}_{\text{RACH}} = n  \}  \Big) 
 = {f^{t-1}_{\text{Prea},0}}  \big(1-\frac{1}{f^{t-1}_{\text{Prea},0}} \big)^{n} .
\end{align}
Due to that the actual number of preambles experiencing idles $ V^{t-1}_{{\rm ip},0}$ can be observed at the eNB, the number of RACH attempting IoT devices in the $(t-1)$th TTI $\zeta ^{t-1} $ can be estimated as
\begin{align}\label{aa3}
\zeta^{t-1}  & =  f^{-1}({\mathbb E} \{ V^{t-1}_{{\rm ip},0} \big| D^{t-1}_{\text{RACH},0}) = \text{log}_{(\frac{{f^{t-1}_{\text{Prea},0}}-1}{f^{t-1}_{\text{Prea},0}})}(\frac{V^{t-1}_{{\rm ip},0}}{f^{t-1}_{\text{Prea},0}}),
\end{align}

To obtain the estimated number of RACH attempting IoT devices in the $t$th TTI $\tilde{D}^t_{\text{RACH},0}$, we also need to know the difference between the estimated numbers of RACH attempting IoT devices in the $(t-1)$th and the $t$th TTIs, denoted by $\delta^{t}$, where $\delta^{t} = \tilde{D}^t_{\text{RACH},0} − \tilde{D}^{t-1}_{\text{RACH},0}$ for $t = 1, 2, \cdots$, and ${\tilde{D}^{0}_{\text{RACH},0}} = 0$. However, $\tilde{D}^{t}_{\text{RACH},0}$ cannot be obtained before the $t$th TTI. To solve this, we can assume $\delta^{t} \approx \delta^{t-1}$ according to \cite{duan2016d}. This is due to that the time between two consecutive TTIs is small, and the available preambles are gradually updated leading to that the number of successful RACH IoT devices does not change significantly in these two TTIs \cite{duan2016d}. Therefore, the number of RACH attempting IoT devices in the $t$th time slot is estimated as
\vspace*{-0.1cm}
\begin{align}\label{1q10}
& {\tilde{D}^t_{\text{RACH},0}}  =  \text{max} \big\{  2V^{t-1}_{{\rm cp},0},  \zeta^{t-1} + \delta^{t-1} \big\} ,
\end{align}
where $2V^{t-1}_{{\rm cp},0}$ represents that there are at least $2V^{t-1}_{{\rm cp},0}$ number of IoT devices colliding in the last TTI.

\subsubsection{Uplink Resource Configuration Based on Load Estimation}
In the following, we propose LE-URC by taking into account the resource condition given in Eq. (\ref{q5-1}). The number of RACH periods $n_{\text{Rach},0}$ and the repetition value $n_{\text{Repe},0}$ is fixed, and only the number of preambles in each RACH period $f_{\text{Prea},0}$ is dynamically configured in each TTI. Using the estimated number of RACH attempting IoT devices in the $t$th TTI ${\tilde{D}^t_{\text{RACH},0}}$, the probability that only one IoT device selects preamble $j$ (i.e., no collision occurs) is expressed as
\vspace*{-0.2cm}
\begin{align}\label{aa4}
{\mathbb P} \{ d_j = 1 \big| {\tilde{D}^t_{\text{RACH},0}} = n  \}=  \big( {\begin{array}{*{20}{c}}
   {\rm{n}}  \\
   {\rm{1}}  \\
\end{array}} \big)
\frac{1}{f^t_{\text{Prea},0}} \big(1-\frac{1}{f^t_{\text{Prea},0}} \big)^{n-1}.
\end{align}
The expected number of RACH attempting IoT devices in the $t$th TTI is derived as
\vspace*{-0.1cm}
\begin{align}\label{aa5}
 & {\mathbb E} \{ {\cal V}^{t}_{{\rm RACH},0} \big| \tilde{D}^t_{\text{RACH},0}  = n  \} = \sum\limits_{j>1}^{f^t_{\text{Prea},0}} {\mathbb P} \{ d_j = 1 \big| \tilde{D}^t_{\text{RACH},0} = n \} = n \big(1-\frac{1}{f^t_{\text{Prea},0}} \big)^{n-1} ,
\end{align}
Based on (\ref{aa5}), the expected number of IoT devices requesting uplink resource in the $t$th TTI is derived as
\vspace*{-0.2cm}
\begin{align}\label{aa51}
{\mathbb E} \{ {\cal V}^{t}_{{\rm reqs}} \big| {\tilde{D}^t_{\text{RACH},0}}= n  \}   &= {\mathbb E} \{ {\cal V}^{t}_{{\rm RACH},0} \big| \tilde{D}^t_{\text{RACH},0}  = n  \} + V^{t}_{\text{un},0}
= n \big(1-\frac{1}{f^t_{\text{Prea},0}} \big)^{n-1} + V^{t-1}_{\text{un},0} ,
\end{align}
where $V^{t-1}_{\text{un},0}$ is the number of unscheduled IoT devices in the last TTI. Note that $V^{t-1}_{\text{un},0}$ can be observed.

However, if the data resource is not sufficient (i.e., occurs when Eq. (\ref{q5-1}) is invalid), some IoT devices may not be scheduled in the $t$th TTI. The upper bound of the number of scheduled IoT devices $ {\cal V}^{t}_{\text{up},0}$ is expressed as
\vspace*{-0.2cm}
\begin{align}\label{1q4}
{\cal V}^{t}_{\text{up},0} =  \frac{R^t_\text{DATA}}{{r}^t_{\text{DATA},i}} = \frac{R_{\rm Uplink} - R^t_{\rm RACH}}{{r}^t_{\text{DATA},i}}  .
\end{align}
where $R_\text{uplink}$ is the total number of REs reserved for uplink transmission in a TTI, $R^t_{\rm RACH}$ is the uplink resource configured for RACH in the $t$th TTI. ${r}^t_{\text{DATA},0}$ is required REs for serving one IoT device given in Eq. (\ref{q5-1}).

According to (\ref{aa51}) and (\ref{1q4}), the expected number of the successfully served IoT devices is given by
\begin{align}\label{1q5}
 {\cal V}^{t}_{suss} ({f^t_{\text{Prea},0}}) & =  \textbf{min} \quad \{{\mathbb E} \{ {\cal V}^{t}_{{\rm reqs}} \big| {\tilde{D}^t_{\text{RACH},0}} = n  \} , {\cal V}^{t}_{\text{up},0}\}.
\end{align}

The maximal expected number of the successfully served IoT devices is obtained by selects the number of preamble $f^{t*}_{\text{Prea},0}$ using
\begin{align}\label{1q5-1}
f^{t*}_{\text{Prea},0}  =  \mathop {\textbf{argmax}}\limits_{ f\in {\boldsymbol{\cal N}}_\text{Prea}} \quad {\cal V}^{t}_{suss} (f) .
\end{align}


The LE-URC approach based on the estimated load ${\tilde{D}^t_{\text{RACH},0}} $ is detailed in \textbf{Algorithm \ref{a3}}. For comparison, we consider an ideal scenario that the actual number of RACH requesting IoT devices $D^t_{\text{RACH}}$ is available at the eNB, namely, Full State Information based URC (FSI-URC). FSI-URC configures $f^{t*}_{\text{Prea},0}$ still using the approach given in Eq. (\ref{1q5-1}), while the load estimation approach given in Section III.B.1) is not required.

\vspace*{-0.2cm}
\begin{algorithm}
\footnotesize
\SetKwData{Left}{left}
\SetKwData{This}{this}
\SetKwData{Up}{up}
\SetKwFunction{Union}{Union}
\SetKwFunction{FindCompress}{FindCompress}
\SetKwInOut{Input}{input}
\SetKwInOut{Output}{output}
\caption{Load Estimation Based Uplink Resource Configuration (LE-URC)} \label{a3}
\Input{The set of the number of preambles in each RACH period ${\boldsymbol{\cal F}}_{\text{Prea},0}$, Number of IoT devices $D$, Operation Iteration $I$.}
\BlankLine
\For{Iteration $\leftarrow 1$ \KwTo $I$}{
Initialization of ${V^{0}_{{\rm ip},0}}:=  12$, ${V^{0}_{{\rm cp},0}}:=  0$, ${\tilde{D}^0_{\text{RACH},0}}  :=  0 $, $\delta^{1} := 0$, and bursty traffic arrival rate $\mu_{bursty}^0=0$\;
\For{$t \leftarrow 1$ \KwTo $T$}{
Generate $\mu_{bursty}^{t}$ using Eq. (\ref{q8})\;
The eNB observes ${V^{t-1}_{{\rm ip},0}}$ and ${V^{t-1}_{{\rm cp},0}}$, and calculate $\zeta^{t-1}$ using Eq. (\ref{aa3})\; 
Estimate the number of RACH requesting IoT devices ${\tilde{D}^t_{\text{RACH},0}}$ using Eq. (\ref{1q10})\;
Select the number of preambles $f^{t*}_{\text{Prea},0}$ using Eq. (\ref{1q5-1}) based on the estimated load ${\tilde{D}^t_{\text{RACH},0}}$\;
The eNB broadcasts $f^{t*}_{\text{Prea},0}$ , and backlogged IoT devices attempt communication in the $t$th TTI\;
Update $\delta^{t+1} := {\tilde{D}^t_{\text{RACH},0}} - {\tilde{D}^{t-1}_{\text{RACH},0}}  $.
}
}
\end{algorithm}

\vspace*{-0.4cm}
\subsubsection{LE-URC for Multiple CE Groups}
\vspace*{-0.2cm}
We slightly revise the introduced single-parameter single-group LE-URC approach (given in Section III.B) to dynamically configure resource for multiple CE groups. Note that the repetition value ${n}_{\text{Repe},i}$ in the LE-URC approach is still defined as a constant to enable the availability of load estimation in Eq. (\ref{1q10}). Remind that the principle of LE-URC approach is to optimize the expectation of the number of successful served IoT devices while balancing $R^t_\text{RACH}$ and $R^t_\text{DATA}$ with limited uplink resource $R_\text{uplink} = R^t_\text{DATA} + R^t_\text{RACH}$. In the multiple CE groups scenarios, the resource $R^t_\text{DATA}$ are allocated to IoT devices in any CE groups without bias, but $R^t_\text{RACH}$ is specifically allocated to each CE group. 

Under this condition, the expected number of successfully served IoT devices $ {\cal V}^{t}_{\text{suss},i}$ given in Eq. (\ref{1q5}) needs to be modified by taking into account multiple variables, which becomes non-convex, and extremely complicates the optimization problem. To solve it, we use a sub-optimal solution by artificially setting uplink resource constrain $R_{\text{Uplink},i}$ for each CE group ($R_\text{Uplink} = \sum_{i=0}^{2} R_{\text{Uplink},i}$). Each CE group can independently allocate the resource between $R^t_{\text{DATA},i}$ and $R^t_{\text{RACH},i}$\hspace*{-0.05cm} according to the approach given in \hspace*{-0.05cm}Eq. \hspace*{-0.05cm}(\ref{1q5-1}).

\vspace*{-0.1cm}
\section{Q-Learning Based Resource Configuration in Single-Parameter Single-Group Scenario}
\vspace*{-0.0cm}

The RL approaches are well-known in addressing dynamic control problem in complex POMDPs \cite{sutton2017reinforcement}. Nevertheless, they have been rarely studied in handling the resource configuration in slotted-Aloha based wireless communication systems. Therefore, it is worthwhile to evaluate the capability of RL in the single-parameter single-group scenario first, in order to be compared with conventional heuristic approaches. In this section, we consider one single CE group with the \textit{fixed RACH periods} $n_{\text{Rach},0}$ as well as the \textit{fixed repetition value} $n_{\text{Repe},0}$, and only dynamically configuring the number of preambles $f_{\text{Prea},0}$ at the beginning of each TTI. In the following, We first study tabular-Q based on the tabular representation of the value function, which is the simplest Q-learning form with guaranteed convergence \cite{sutton2017reinforcement}, but requires extremely long training time. We then study Q-learning with function approximators to improve training efficiency, where LA-Q and DQN will be used to construct an approximation of the desired value function.

\vspace*{-0.2cm}
\subsection{Q-Learning and Tabular Value Function}
\vspace*{-0.1cm}

Considering a Q-agent deployed at the eNB to optimize the number of successfully served IoT devices in real-time, the Q-agent need to explore the environment in order to choose appropriate actions progressively leading to the optimization goal. We define $s \in \cal S$, $a \in \cal A$, and $r \in \cal R$ as any state, action, and reward from their corresponding sets, respectively. At the beginning of the $t$th TTI ($t \in \{0, 1, 2, \cdots\}$), the Q-agent first observes the current state $S^t$ corresponding to a set of previous observations ($O^t$=$\{U^{t-1}, U^{t-2}, \cdots,U^{1}$\}) in order to select an specific action $A^t \in {\cal A}(S^t)$. The action $A^t$ corresponds to the number of preambles in each RACH period $f^t_{\text{Prea},0}$ in single CE group scenario.

\vspace*{-0.1cm}
\captionsetup{singlelinecheck=false}  
\begin{figure}[htbp!]
\setlength{\abovecaptionskip}{0pt}
    \begin{center}
        \includegraphics[width=1\textwidth]{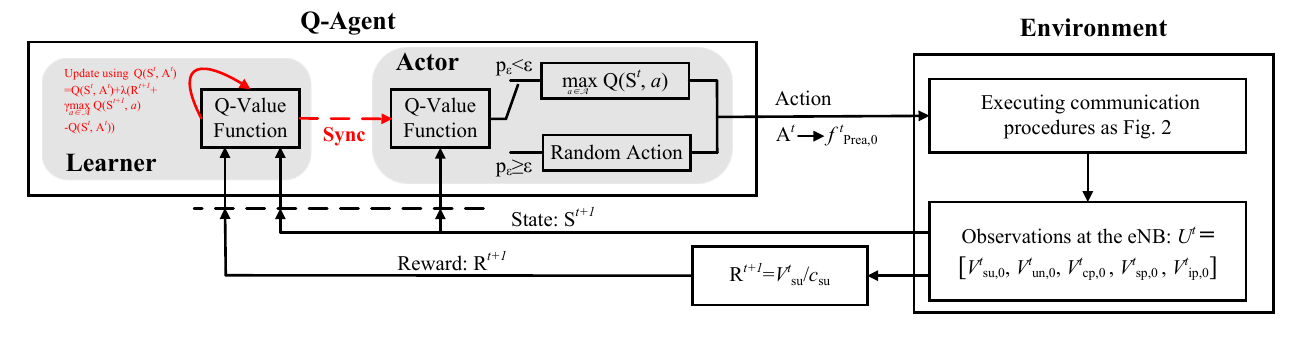}
        \caption{The Tabular-Q agent and environment interaction in the POMDP.
        }\label{fig:q}
    \end{center}
\end{figure}
\vspace*{-0.0cm}

As shown in Fig. \ref{fig:q}, we consider a basic state function in the single CE group scenario, where $S^{t}$ is a set of indices mapping to the current observed information $U^{t-1} = [ V^{t-1}_{{\rm su},0} , V^{t-1}_{{\rm un},0}, V^{t-1}_{{\rm cp},0}, V^{t-1}_{{\rm sp},0}, V^{t-1}_{{\rm ip},0} ]$. With the knowledge of the state $S^{t}$, the Q-agent chooses an action $A^t$ from the set ${\cal A}$, which is a set of indexes mapped to the set of the number of available preambles ${\boldsymbol{\cal F}}_{\text{Prea}}$. Once an action $A^t$ is performed, the Q-agent will receive a scalar reward $R^{t+1}$, and observe a new state $S^{t+1}$. The reward $R^{t+1}$ indicates to what extent the executed action $A^t$ can achieve the optimization goal, which is determined by the new observed state $S^{t+1}$. As the optimization goal is to maximize the number of the successfully served IoT devices, we define the reward $R^{t+1}$ as a function that positively proportional to the observed number of successfully served IoT devices $V^{t}_{{\rm su}} \in O^{t}$, which is defined as
\vspace*{-0.2cm}
\begin{align}\label{q12}
R^{t+1}  =  V_{{\rm su}}^{t} / c_{{\rm su}},
\end{align}
where $c_{{\rm su}}$ is constant used to normalize the reward function.

Q-learning is a value-based RL approach \cite{sutton2017reinforcement,mnih2015human}, where the policy of states to actions mapping $\pi(s) = a$ is learned using a state-action value function $Q(s, a)$ to determine an action for the state $s$. We first use a \textit{lookup table} to represent the state-action value function $Q(s, a)$ (tabular-Q), which consists of value scalars for all the state and action spaces. To obtain an action $A^t$, we select the highest value scalar from the numerical value vector $Q(S^t, a)$, which maps all possible actions under $S^t$ to the Q-value table $Q(s, a)$.

Accordingly, our objective is to find an optimal Q-value table $Q^*(s, a)$ with optimal policy $\pi^*$ that can select actions to dynamically optimize the number of served IoT devices. To do so, we train a initial Q-value table $Q(s, a)$ in the environment using Q-Learning algorithm, where $Q(s, a)$ is immediately updated using the current observed reward ${\cal R}^{t+1} $ after each action as
\vspace*{-0.23cm}
\begin{align}\label{q11}
Q(S^{t}, A^{t})  = &  Q(S^{t}, A^{t}) +  \lambda \big[ R^{t+1} + \gamma \mathop {\text{max}}\limits_{a\in {\cal A}} Q(S^{t+1}, a) - Q(S^{t}, A^{t}) \big] ,
\end{align}
where $\lambda$ is a constant step-size learning rate that affects how fast the algorithm adapt to a new environment, $\gamma \in [0,1)$ is the discount rate that determines how current rewards affects the value function updating, $\mathop{\text{max}}\limits_{a\in {\cal A}} Q(S^{t+1}, a)$ approximates the value in optimal Q-value table $Q^*(s, a)$ via the up-to-date Q-value table $Q(s, a)$ and the obtained new state $S^{t+1}$. Note that $Q(S^{t}, A^{t})$ in Eq. (\ref{q11}) is a scalar, which means that we can only update one value scalar in the Q-value table $Q(s, a)$ with one received reward $R^{t+1}$.


As shown in Fig. \ref{fig:q}, we consider $\epsilon$-greedy approach to balance \textit{exploitation} and \textit{exploration} in the Actor of the Q-Agent, where $\epsilon$ is a positive real number and $\epsilon \le 1$. In each TTI $t$, the Q-agent randomly generates a probability $p^t_{\epsilon}$ to compare with $\epsilon$. Then, with the probability $\epsilon$, the algorithm randomly chooses an action from the remaining feasible actions to improve the estimate of the non-greedy action's value. With the probability $1-\epsilon$, the algorithm exploits the current knowledge of the Q-value table to choose the action that maximizes the expected reward. 

Particularly, the learning rate $\lambda$ is suggested to be set to a small number (e.g., $\lambda=0.01$) to guarantee the stable convergence of Q-value table in this NB-IoT communication system. This is due to that a single reward in a specific TTI can be severely biased, because state function is composed of multiple unobserved information with unpredictable distributions (e.g., an action allows for the setting with large number of preambles $f^t_{\text{prea}}$, but massive random collisions accidentally occur, which leads to an unusual low reward). In the following, the implementation of uplink resource configuration using tabular-Q based real-time optimization is shown in \textbf{Algorithm \ref{a1}}.

\vspace*{-0.0cm}
\begin{algorithm}
\footnotesize
\SetKwData{Left}{left}
\SetKwData{This}{this}
\SetKwData{Up}{up}
\SetKwFunction{Union}{Union}
\SetKwFunction{FindCompress}{FindCompress}
\SetKwInOut{Input}{input}
\SetKwInOut{Output}{output}
\caption{Tabular-Q Based Uplink Resource Configuration} \label{a1}
\Input{Valid numbers of preambles ${\boldsymbol{\cal F}}_{\text{Prea}}$, Number of IoT devices $D$, Operation Iteration $I$.}
\BlankLine 
Algorithm hyperparameters: learning rate $\lambda \in (0,1]$, discount rate $\gamma \in [0,1)$, $\epsilon$-greedy rate $\epsilon \in (0,1]$ \;
Initialization of the Q-value table $Q(s, a)$ with $0$ value scalars\;
\For{Iteration $\leftarrow 1$ \KwTo $I$}{
Initialization of $S^1$ by executing a random action $A^0$ and bursty traffic arrival rate $\mu_{bursty}^0=0$\;
\For{$t \leftarrow 1$ \KwTo $T$}{
Update $\mu_{bursty}^{t}$ using Eq. (\ref{q8})\;
\lIf{$p^t_{\epsilon}<\epsilon$}{select a random action $A^t$ from $\cal A$}
\lElse{select $A^t= \mathop {\text{argmax}}\limits_{{a\in {\cal A}}} Q(S^t, a)$ }
The eNB broadcasts ${f^t_\text{Prea}}={\boldsymbol{\cal F}_\text{Prea}}(A^t)$ and backlogged IoT devices attempt communication in the $t$th TTI\;
The eNB observes $S^{t+1}$, calculate the related $R^{t+1}$ using Eq. (\ref{q12}), and update $Q({ S^t}, { A^t})$ using Eq. (\ref{q11}).
}
}
\end{algorithm}
\vspace*{-0.0cm}

\vspace*{-0.5cm}
\subsection{Value Function Approximation}
\vspace*{-0.1cm}
Since tabular-Q needs its each element to be updated to converge, searching for an optimal policy can be difficult in limited time and computational resource. To solve this problem, we use a value function approximator instead of Q-value table to find a sub-optimal approximated policy. Generally, selecting a efficient approximation approach to represent the value function for different learning scenarios is a usual problem within the RL \cite{sutton2017reinforcement,konidaris2011value,thrun1993issues,hauskrecht2000value}. A variety of function approximation approaches can be conducted, such as LA, DNNs, tree search, and which approach to be selected can critically influence the successful learning \cite{sutton2017reinforcement,thrun1993issues,hauskrecht2000value}. The function approximation should fit the complexity of the desired value function, and be efficient to obtain good solutions. Unfortunately, most function approximation approaches require specific design for different learning problems, and there is no basis function, which is both reliable and efficient to satisfy all learning problems. 

In this subsection, we first focus on the linear function approximation for Q-learning, due to its simplicity, efficiency, and guaranteed convergence \cite{sutton2017reinforcement,geramifard2013tutorial,melo2007q}. We then conduct the DNN for Q-learning as a more effective but complicated function approximator, which is also known as DQN \cite{mnih2015human}. The reasons we conduct DQN are that: 1) the DNN function approximation is able to deal with several kinds of partially observable problems \cite{sutton2017reinforcement,mnih2015human}; 2) DQN has the potential to accurately approximate the desired value function while addressing a problem with very large state spaces \cite{mnih2015human}, which can be favored for the learning in the multiple CE group scenarios; 3) DQN is with high scalability, where the scale of its value function can be easily fit to a more complicated problem; 4) a variety of libraries have been established to facilitate building DNN architectures and accelerate experiments, such as TensorFlow, Pytorch, Theano, Keras, and etc..

\subsubsection{Linear Approximation}
LA-Q uses a linear weight matrix $\bf w$ to approximate the value function $Q(s, a)$ with feature vector ${\vec x} = {\bf x}(s)$ corresponding to the state $S^t$. The dimensions of weight matrix $\bf w$ is $ |{\cal A}| \times |{\vec x}|$, where $|{\cal A}|$ is the total number of all available actions and $|{\vec x}|$ is the size of feature vector ${\vec x}$. Here, we consider polynomial regression (as \cite[Eq. 9.17]{sutton2017reinforcement}) to construct the real-valued feature vector ${\bf x}(s)$ due to its efficiency\footnote{The polynomial case is the most well understood feature constructor and always performs well in practice with appropriate setting \cite{sutton2017reinforcement,konidaris2011value}. Furthermore, the results in \cite{cheng2018polynomial} shows that there is a rough correspondence between a fitted neural network and a fitted ordinary parametric polynomial regression model. These reasons encourage us to compare the polynomial based LA-Q with DQN}. In the training process, the exploration is the same as the tabular Q-learning by generating random actions, but the exploitation is calculated using the weight matrix $\bf w$ of the value function. In detail, to predict an action using the LA value function $Q(S^t, a, \bf w)$ with state $S^t$ in the $t$th TTI, the approximated value function scalars for each action ${\bf a}$ is obtained by inner-producting between the weight matrix $\bf w$ and the features vector ${\bf x}(s)$ as:
\vspace*{-0.13cm}
\begin{align}\label{q15}
& Q(S^t, a, {\bf w})  =  {\bf w} \cdot  {\bf x}(S^t)^T  =  \Big[ \sum_{j=0}^{|{\vec x}|-1} w_{(0,j)} x_j(S^t), \sum_{j=0}^{|{\vec x}|-1} w_{(1,j)} x_j(S^t), \cdots ,\sum_{j=0}^{|{\vec x}|-1} w_{(|{\cal A}|-1,j)} x_j(S^t) \Big]^T .
\end{align}
By searching for the maximal value function scalar in $Q(S^t, a, {\bf w})$ given in Eq. (\ref{q15}), we can obtain the matched action $A^{t}$ to maximize future rewards.

To obtain the optimal policy, we update the weigh matrix $\bf w$ in the value function $Q(s, a; \bf w)$ using Stochastic Gradient Descent (SGD) \cite{sutton2017reinforcement,bishop2006pattern}. SGD minimizes the error on predictions of observation after each example, where the error is reduced by a small amount following the direction to the optimal target policy $Q^*(s,a)$. As it is infeasible to obtain optimal target policy by summing over all states, we instead estimate the desired action-value function by simply considering one learning sample $Q^*(s,a) \approx Q^*(S^t,a, {\bf w}^t)$ \cite{sutton2017reinforcement}. In each TTI, the weigh matrix $\bf w$ is updated following
\vspace*{-0.23cm}
\begin{align}\label{q16}
& {\bf w}^{t+1} =  {\bf w}^t  -   \lambda \nabla L({\bf w}^t ),
\end{align}
where $\lambda$ is the learning rate. $\nabla L({\bf w}^t )$ is the gradient of the loss function $L({\bf w}^t)$ used to train the Q-function approximator. This is given as
\vspace*{-0.2cm}
\begin{align}\label{q17}
&  \nabla L({\bf w}^t )  = \big( R^{t+1} + \gamma \mathop {\text{max}}\limits_{{a}} Q(S^{t+1}, a; {\bf w}^t) -  Q(S^t,a, {\bf w}^t) \cdot  {\bf x}(A^t,S^t)^T \big)\cdot  \nabla_{\bf w}Q(S^t,A^t, {\bf w}^t)
\end{align}
where ${\bf w}^t$ is the weight matrix, ${\bf x}(A^t,S^t)$ is the features matrix with the same shape of ${\bf w}^t$. ${\bf x}(A^t,S^t)$ is constructed by zeros and the feature vector located in the row corresponding to the index of the action selected in the $t$th TTI $A_t$. Note that $ Q(S^{t+1},a;{\bf w}^t)$ is a scalar. The learning procedure follows \textbf{Algorithm \ref{a1}} by changing the Q-table $Q(s, a)$ to the LA value function $Q(s, a; \bf w)$ with linear weigh matrix $\bf w$, and updating $Q(s, a; \bf w)$ with SGD given in (\ref{q17}) in step 10 of \textbf{Algorithm \ref{a1}}.

\subsubsection{Deep Q-Network}

The DQN agent parameterizes the action-state value function $Q(s,a)$ by using a function $Q(s, a; \bm{\theta})$, where $\bm{\theta}$ represents the weights matrix of a DNN with multiple layers. We consider the conventional DNN, where neurons between two adjacent layers are fully pairwise connected, namely fully-connected layers. The input of the DNN is given by the variables in state $S^t$; the intermediate hidden layers are Rectifier Linear Units (ReLUs) by using the function $ f(x)= \text{ max }(0,x)$; while the output layer is composed of linear units\footnote{
Linear activation is used here according to \cite{mnih2015human}. Note that Q-learning is value-based, thus the desired value function given in Eq. (15) can be bigger than 1, rather than a probability, and thus the activation function with return value limited in $[−1,1]$ (such as sigmoid function and tanh function) can lead to convergence difficulty.}, which are in one-to-one correspondence with all available actions in $\cal A$.

\vspace*{-0.1cm}
\captionsetup{singlelinecheck=false}  
\begin{figure}[htbp!]
\setlength{\abovecaptionskip}{0pt}
    \begin{center}
        \includegraphics[width=1\textwidth]{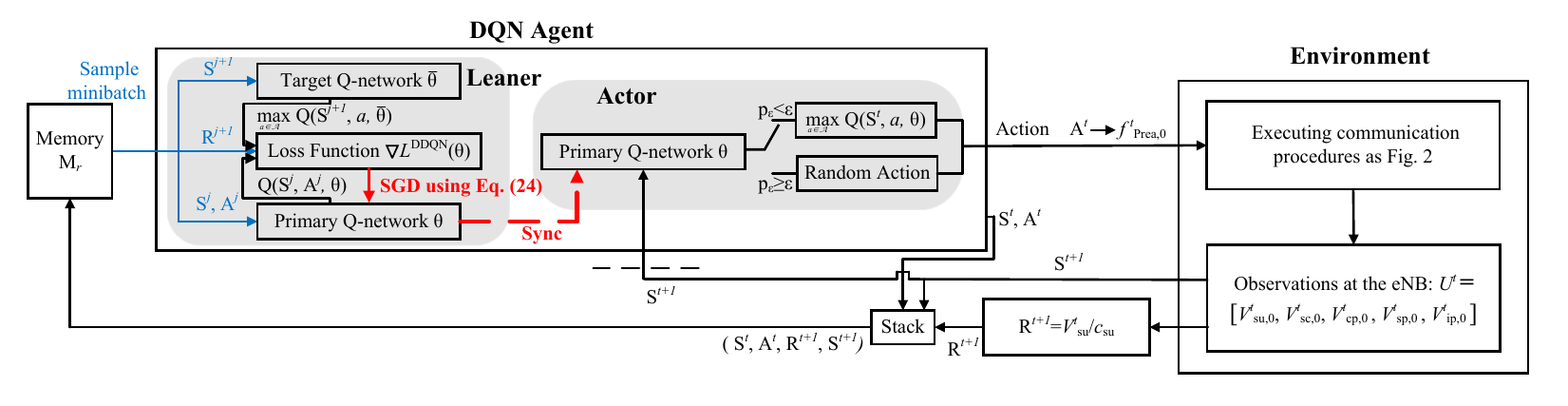}
        \caption{The DQN agent and environment interaction in the POMDP.
        }\label{fig:dnnq}
    \end{center}
    \vspace*{-0.8cm}
\end{figure}

\begin{algorithm}
\footnotesize
\SetKwData{Left}{left}
\SetKwData{This}{this}
\SetKwData{Up}{up}
\SetKwFunction{Union}{Union}
\SetKwFunction{FindCompress}{FindCompress}
\SetKwInOut{Input}{input}
\SetKwInOut{Output}{output}
\caption{DQN Based Uplink Resource Configuration} \label{a2}
\Input{The set of numbers of preambles in each RACH period ${\boldsymbol{\cal F}}_{\text{Prea}}$, the number of IoT devices $D$, and operation iteration $I$.}
\BlankLine 
Algorithm hyperparameters: learning rate $\lambda \in (0,1]$, discount rate $\gamma \in [0,1)$, $\epsilon$-greedy rate $\epsilon \in (0,1]$, target network update frequency $K$\;
Initialization of replay memory $M$ to capacity $C$, the primary Q-network ${ \boldsymbol \theta}$, and the target Q-network $\bar{\bm{\theta}}$\;
\For{Iteration $\leftarrow 1$ \KwTo $I$}{
Initialization of $S^1$ by executing a random action $A^0$ and bursty traffic arrival rate $\mu_{bursty}^0=0$\;
\For{$t \leftarrow 1$ \KwTo $T$}{
Update $\mu_{bursty}^{t}$ using Eq. (\ref{q8})\;
\lIf{$p_{\epsilon}<\epsilon$}{select a random action $A^t$ from $\cal A$}
\lElse{select  $A^t=\mathop {\text{argmax}}\limits_{{a\in {\cal A}}} Q(S^t, a, { \boldsymbol \theta})$}
The eNB broadcasts ${\boldsymbol{\cal F}}_{\text{Prea}}(A^t)$ and backlogged IoT devices attempt communication in the $t$th TTI\;
The eNB observes $S^{t+1}$, and calculate the related $R^{t+1}$ using Eq. (\ref{q12})\;
Store transition $(S^{t}, A^t, R^{t+1}, S^{t+1})$ in replay memory $M$\;
Sample random minibatch of transitions $(S^{j}, A^j, R^{j+1}, S^{j+1})$ from replay memory $M$\;
Perform a gradient descent for $Q(s, a; { \boldsymbol \theta})$ using Eq. (\ref{q19})\;
Every $K$ steps update target Q-network $\bar{\bm{\theta}}= {\boldsymbol \theta}$.
}
}
\end{algorithm}

The exploitation is obtained by performing forward propagation of Q-function $Q(s, a; \bm{\theta})$ with respect to the observed state $S^t$. The weights matrix $\bm{\theta}$ is updated online along each training episode by using double deep Q-learning (DDQN) \cite{van2016deep}, which to some extend reduce the substantial overestimations\footnote{Overestimation refers to that some suboptimal actions regularly were given higher Q-values than optimal actions, which can negatively influence the convergence capability and training efficiency of the algorithm \cite{van2016deep,thrun1993issues}.} of value function. Accordingly, learning takes place over multiple training episodes, with each episode of duration $N_\text{TTI}$ TTI periods. In each TTI, the parameter $\bm{\theta}$ of the Q-function approximator $Q(s,a;\bm{\theta})$ is updated using SGD as
\begin{align}\label{q19-2}
{\boldsymbol \theta}^{t+1} =  {\boldsymbol \theta}^{t}  -   \lambda_\text{RMS} \nabla L^\text{DDQN}({\boldsymbol \theta}^{t}),
\end{align}
where $\lambda_\text{RMS}$ is RMSProp learning rate \cite{tieleman2012lecture}, $\nabla L(\theta)$ is the gradient of the loss function $L(\theta^t)$ used to train the Q-function approximator. This is given as
\begin{align}\label{q19}
 \nabla L^\text{DDQN}(\theta^t) =  &  {\mathbb E}_{S^i,A^i,R^{i+1},S^{i+1}} \big[\big( R^{i+1} + \gamma  \mathop{\text{max}}\limits_{{a}}Q(S^{i+1}, a; \bar{\bm{\theta}}^t) -  Q(S^i, A^i;   \bm{\theta}^t) \big) \nabla_{\bm{\theta}} Q(S^i, A^i; \bm{\theta}^t)\big],
\end{align} 
where the expectation is taken with respect to a so-called minibatch, which are randomly selected previous samples $(S^i,A^i,S^{i+1},R^{i+1})$ for some $i\in\{t-M_r,...,t\}$, with $M_r$ being the replay memory \cite{mnih2015human}. When $t-M_r$ is negative, this is interpreted as including samples from the previous episode. The use of minibatch, instead of a single sample, to update the value function $Q(s, a; {\boldsymbol \theta})$ improves the convergent reliability of value function \cite{mnih2015human}. Furthermore, following DDQN \cite{van2016deep}, in (\ref{q19}), $\bar{\bm{\theta}}^t$ is a so-called target Q-network that is used to estimate the future value of the Q-function in the update rule. This parameter is periodically copied from the current value $\bm{\theta}^t$ and kept fixed for a number of episodes \cite{van2016deep}.

\vspace*{-0.1cm}
\section{Q-Learning Based Resource Configuration in Multi-Parameter Multi-Group Scenario}
\vspace*{-0.0cm}
Practically, NB-IoT is always deployed with multiple CE groups to serve IoT devices with various coverage requirements. In this section, we study the problem (\ref{q6-1}) of optimizing the resource configuration for three CE groups each with parameters $A^t=\{n^t_{\text{Rach},i},f^t_{\text{Prea},i},$ $n^t_{\text{Repe},i}\}_{i=0}^{2}$. This joint optimization by configuring each parameter in each CE group can improve the overall data access and transmission performance. Note that each CE group shares the uplink resource in the same bandwidth, and the eNB schedules data resource to all RRC connected IoT devices without the CE group bias as introduced in Sec. II.B.4). To optimize the number of served IoT devices in real-time, the eNB should not only balance the uplink resource between RACH and data, but also balance them among each CE group.

The Q-learning algorithms with the single CE group provided in Sec. IV are model-free, and thus their learning structure can be directly used in this multi-parameter multi-group scenario. However, considering multiple CE groups results in the increment of observations space, which exponentially increases the size of state space. To train Q-agent with this expansion, the requirements of time and computational resource greatly increase. In such case, the tabular-Q would be extremely inefficient, as not only the state-action value table requires a big memory, but it is impossible to repeatedly experience every state to achieve convergence with limited time. In view of this, we only study Q-learning with value function approximation (LA-Q and DQN) to design uplink resource configuration approaches for the multi-parameter multi-group scenario.

LA-Q and DQN are with high capability to handle massive state spaces, and thus we can considerably improve the state spaces with more observed information to support the optimization of Q-agent. Here, we define the current state $S^t$ includes information about the last $M_o$ TTIs ($U^{t-1}, U^{t-2}, U^{t-3}, \cdots, U^{t-M_\text{o}}$). This design improves Q-agent by enabling it to estimate the trend of traffic. As our goal is to optimize the number of served IoT devices, the reward function should be defined according to the number of successfully served IoT devices $V_{{\rm su},i}$ of each CE group, which is expressed as
\begin{align}\label{2q1}
R^{t+1}   =  \big( \sum_{i=0}^{2}  V^t_{{\rm su},i} \big) / c_{{\rm su}}.
\end{align}

Same as the state spaces, the available action spaces also exponentially increases with the increment of the adjustable configurations. The number of available actions corresponds to the possible combinations of configurations
$ |{\cal A}| = \prod\limits_{i=0}^{2}  ( { |{\boldsymbol{\cal N}}_{\text{Rach},i}| \times |{\boldsymbol{\cal N}}_{\text{Repe},i}| \times |{\boldsymbol{\cal F}}_{\text{Prea},i}}| ) $ (i.e., $|\boldsymbol \cdot|$ denotes the number of elements in any vector $\boldsymbol \cdot$, $\cal A$ is the set of actions, ${\boldsymbol{\cal N}}_{\text{Rach},i}$, ${\boldsymbol{\cal N}}_{\text{Repe},i}$, and ${\boldsymbol{\cal F}}_{\text{Prea},i}$ are the sets of the number of RACH periods, the repetition value, and the number of preambles in each RACH period). Unfortunately, it is extremely hard to optimize the system under such numerous action spaces (i.e., $|{\cal A}| $ can be over fifty thousands.), due to that the system will fall into updating policy with only a small part of the action in $\cal A$, and finally leads to convergence difficulty. To solve this problem, we then provide two approaches that can reduce the dimension of action space to enable the LA and DQN in the multi-parameter multi-group scenario.

\subsection{Actions Aggregated Approach}
We first provide AA based Q-learning approaches, which guarantee convergent capability by sacrificing the accuracy of action selection\footnote{The action aggregation has been rarely evaluated, but the same idea, namely, state aggregation has been well studied, which is a basic function approximation approach \cite{sutton2017reinforcement}.}. In detail, the specific action selection can be converted to the increasing or decreasing trend selection. Instead of selecting the exact values from the sets of ${\boldsymbol{\cal N}}_{\text{Rach},i}$, ${\boldsymbol{\cal N}}_{\text{Repe},i}$, and ${\boldsymbol{\cal F}}_{\text{Prea},i}$, we convert it to single step ascent/descent based on the last action, which is represented by $A_{\text{Rach},i}^t \in \{0, 1\}$, $A_{\text{Repe},i}^t \in \{0, 1\}$, and $A_{\text{Prea},i}^t \in \{0, 1\}$ for the number of RACH periods $n^t_{\text{Rach},i}$, the repetition values $n^t_{\text{Repe},i}$, and the number of preambles in each RACH period $f^t_{\text{Prea},i}$ in the $t$th TTI. Consequently, the size of total action spaces for the three CE groups is reduced to $|{\cal A} |$=$2^9$=$512$. By doing so, the algorithms for training with LA function approximator and DQN in the multiple configurations multiple CE groups scenario can be deployed following \textbf{Algorithm \ref{a1}} and \textbf{Algorithm \ref{a2}}, respectively.

\subsection{Cooperative Multi-agent Learning Approach}

\begin{figure}[htbp!]
\setlength{\abovecaptionskip}{0pt}
    \begin{center}
        \includegraphics[width=1\textwidth]{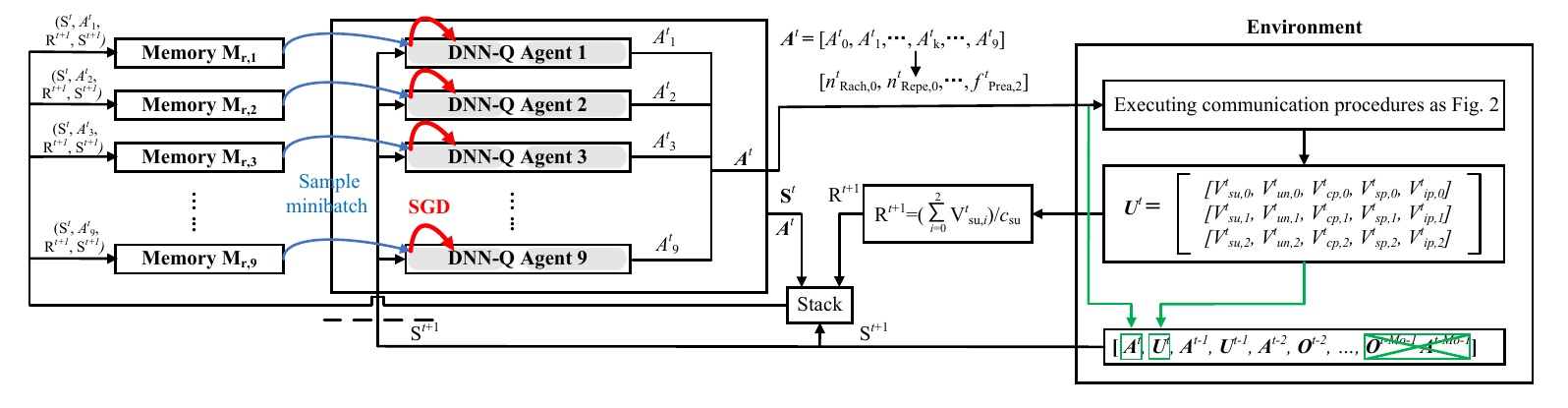}
        \caption{The CMA-DQN agents and environment interaction in the POMDP.
        }\label{fig:cadnnq}
    \end{center}
    \vspace*{-0.8cm}
\end{figure}

Despite that the uplink resource configuration is managed by a central authority, identifying the control of each parameter as one sub-task that is cooperatively handled by independent Q-agents is sufficient to deal with the problem with unsolvable action spaces \cite{busoniu2008comprehensive}. As shown in Fig. \ref{fig:cadnnq}, we consider multiple DQN agents are centralized at the eNB with the same structure of value function approximator\footnote{The structures of value function approximator can also be specifically designed for RL agents with sub-tasks of significantly different complexity. However, there is no such requirement in our problem, so it will not be considered.} following Section IV.B.2). We break down the action space by considering nine separate action variables in $A^t$, where each DQN agent controls their own action variable as shown in Fig. \ref{fig:cadnnq}. Recall that we have three variables for each group $i$, namely ${n}_{\text{Rach},i}$, ${n}_{\text{Repe},i}$, and ${f}_{\text{Prea},i}$. 

\vspace*{-0.1cm}
\captionsetup{singlelinecheck=false}  

We introduce a separate DQN agent for each output variable in $A^t$ defined as action $A^t_k$ selected by the $k$th agent, where each $k$th agent is responsible to update the value $Q(S^t,A_k^t; \bm{\theta}_k)$ of action $A_k^t$ in shared state $S^t$. The DQN agents are trained in parallel and receive the same reward signal given in Eq. (\ref{2q1}) at the end of each TTI as per problem (\ref{q6-1}). The use of this common reward signal ensures that all DQN agents aim at cooperatively increase the objective in (\ref{q6-1}). Note that the approach can be interpreted as applying a factorization of the overall value function akin to the approach proposed in \cite{Sunehag2018} for multi-agent systems.

The challenge of this approach is how to evaluate each action according to the common reward function. For each DQN agent, the received reward is corrupted by massive noise, where its own effect on the reward is deeply hidden in the effects of all other DQN agents. For instance, a positive action can receive a mismatched low reward due to other DQN agents' negative actions. Fortunately, in our scenario, all DQN agents are centralized at the eNB, which means that all DQN agents can have full information among each other. Accordingly, we adopt the action selection histories of each DQN agent as part of state function\footnote{The state function can be designed to collect more information according to the complexity requirements, such as sharing the value function between each DQN agent \cite{busoniu2008comprehensive}.}, thus they are able to know how reward is influenced by different combinations of actions. To do so, we define state variable $S^t$ as
\vspace*{-0.3cm}
\begin{align}\label{2q4}
S^t=[A^{t-1},U^{t-1},A^{t-2},U^{t-2},\cdots,A^{t-M_o},U^{t-M_o}],
\end{align}
where $M_\text{o}$ is the number of stored observations, $A^{t-1}$ is the set of selected action of each DQN agent in the $(t-1)$th TTI corresponding to ${n}_{\text{Rach},i}$, ${n}_{\text{Repe},i}$, and ${f}_{\text{Prea},i}$ for the $i$th CE group, and $U^{t-1}$ is the set of observed transmission receptions. 

In each TTI, the parameters $\theta_k$ of the Q-function approximator $Q(S^t,A^t_k;\theta_k)$ are updated using SGD at all agents $k$ as Eq. (\ref{q19-2}). The learning algorithm can be implemented following \textbf{Algorithm \ref{a2}}. Different from the single-parameter single-group scenario, we need to first initialize nine primary networks ${ \boldsymbol \theta}_{k}$, target networks $\bar{{ \boldsymbol \theta}}_{k}$, and replay memories $M_k$ for each DQN agent. In step 11 of \textbf{Algorithm \ref{a2}}, the current transactions of each DQN agent should be stored in their own memory separately. In step 12 and 13 of \textbf{Algorithm \ref{a2}}, the minibatch of transaction should separately sampled from each memory to train the corresponding DQN agent.

\vspace*{-0.2cm}
\section{Simulation Results}
\vspace*{-0.1cm}

In this section, we evaluate the performance of the proposed Q-learning approaches and compare it with the conventional LE-URC and FSI-URC described in Sec. III via numerical experiments. We adopt the standard network parameters listed in Table \ref{table_1} following \cite{RohdeSchwarz2016white,wang2017primer,3GPP2015Cellular,3GPP2017MAC,3GPP2017PhyCM}, and hyperparameters for Q-learning listed in Table \ref{table_2}. Accordingly, one epoch consists of 937 TTIs (i.e., 10 minutes). The RL agents will first be trained in a so-called learning phase, and after convergence, their performance will be compared with LE-URC and FSI-URC in a so-called testing phase. All testing performance results are obtained by averaging over 1000 episodes. In the following, we present our simulation results of the single-parameter single-group scenario and the multi-parameter multi-group scenario in Section VI-A and Section VI-B, respectively.

\captionsetup{singlelinecheck=true}
\begin{table}[htbp!]
\vspace*{-0.1cm}
	\centering
	\caption{Simulation Parameters}
	\vspace*{-0.3cm}
	{\renewcommand{\arraystretch}{0.6}
		\begin{tabular}{|*{1}{p{6cm}}|*{1}{p{2.11cm}} |*{1}{p{6cm}}|*{1}{p{2.31cm}|} }
			\hline
			\rowcolor{Gray}
		   \bf{Parameters}   &    \bf{Setting}$\vphantom{\Big(}$ &  \bf{Parameters}   &    \bf{Setting}$\vphantom{\Big(}$ \\  \hline 
            Path-loss exponent $\eta$ $\vphantom{\big(}$ & 4 
            & noise power $\sigma^2  $ $\vphantom{\Big(}$ &   -138 dBm  \\
            eNB broadcast power $P_{\text{NPBCH}}$ $\vphantom{\big(}$	&  35 dBm 
            &  Path-loss inverse power control threshold $\rho$ $\vphantom{\big(}$  & 120 dB   \\
            Maximal preamble transmit power $P_{\text{RACHmax}}\vphantom{\big(}$ &  23 dBm  
		    & The received SNR threshold  $\gamma_{\rm th}$ $\vphantom{\big(}$ & 0 dB  \\
          Duration of periodic traffic $T_\text{periodic}$ $\vphantom{\big(}$ & 1 hour
            &   TTI $\vphantom{\big(}$ & $640$ms \\
          Duration of bursty traffic $T_\text{bursty}$ $\vphantom{\big(}$ & 10 minutes 
		    & Set of number of preambles ${\cal F}_\text{Prea}$ $\vphantom{\big(}$ & \{$12,24,36,48$\}   \\
		 Maximum allowed resource requests $\gamma_\text{RRC}$ $\vphantom{\big(}$  & 5 
			&  Set of repetition value ${\cal N}_\text{Repe}$ $\vphantom{\big(}$ &  \{$1,2,4,8,16,32$\} \\ 
	      Maximum RACH attempts $\gamma_{\text{pMax}}$ $\vphantom{\big(}$  & 10
            & Set of number of RACH periods ${\cal N}_\text{Rach}$ $\vphantom{\big(}$ & \{$1,2,4$\}  \\
	      Maximum allowed RACH in one CE $\gamma_{\text{pCE},i}$ $\vphantom{\big(}$ & 5 & REs required for $B_\text{RACH}$ & 4 \\
          Bursty traffic parameter Beta(${\alpha},{\beta}$) &  (3,4)
           & REs required for $B_\text{DATA}$ & 32  \\
             \hline
		\end{tabular}
	}
	\label{table_1}
\end{table}

\captionsetup{singlelinecheck=true}
\begin{table}[htbp!]
\vspace*{-0.2cm}
	\centering
	\caption{Q-learning Hyperparameters}
	\vspace*{-0.3cm}
	{\renewcommand{\arraystretch}{0.6}
		\begin{tabular}{|*{1}{p{6cm}}|*{1}{p{2.11cm}} |*{1}{p{6cm}}|*{1}{p{2.31cm}|} }
			\hline
			\rowcolor{Gray}
		   \bf{Hyperparameters}   &    \bf{Value}$\vphantom{\Big(}$ &  \bf{Hyperparameters}   &    \bf{Value}$\vphantom{\Big(}$ \\  \hline
		   $\vphantom{\big(}$Learning rate $\lambda$ for Tabular-Q and LA-Q  & 0.01
		   & Learning rate by RMSProp $\lambda_\text{RMS}$ for DQN  & 0.0001 \\
			$\vphantom{\big(}$Initial	exploration $\epsilon$  & 1
			& Final	exploration $\epsilon$  & 0.1 \\
		   $\vphantom{\big(}$Discount rate $\gamma$ & 0.5 
           & Minibatch size $\vphantom{\big(}$ & 32 \\
           $\vphantom{\big(}$Replay memory & 10000
           & Target Q-network update frequency & 1000 \\
             \hline
		\end{tabular}
	}
	\label{table_2}
	\vspace*{-0.3cm}
\end{table}

\vspace*{-0.4cm}
\subsection{Single-Parameter Single-Group Scenario}
\vspace*{-0.0cm}
In the single-parameter single-group scenario, eNB is located at the center of a circular area with a 10 km radius, and the IoT devices are randomly located within the cell. We set the number of RACH periods as ${n}_\text{Rach}=1$, the repetition value as ${n}_\text{Repe}=4$, and the limited uplink resource as $R_\text{uplink}=1536$ REs (i.e., 32 slots with 48 sub-carriers). Unless otherwise stated, we consider the number of periodical IoT devices to be $D_\text{periodic}=10000$, and the number of bursty IoT devices to be $D_\text{bursty}=5000$. The DQN is set with three hidden layers, each with 128 ReLU units. Tabular-Q, LA-Q, and DQN approaches are proposed in Sec. IV.A, IV.B.1), and IV.B.2), respectively. The conventional LE-URC and FSI-URC approaches are proposed in Sec. III.B.

\captionsetup{singlelinecheck=false}  
\begin{figure}[htbp!]
\vspace*{-0.6cm}
    \begin{center}
    \begin{minipage}[t]{0.44\textwidth}
    \centering
        \includegraphics[width=1\textwidth]{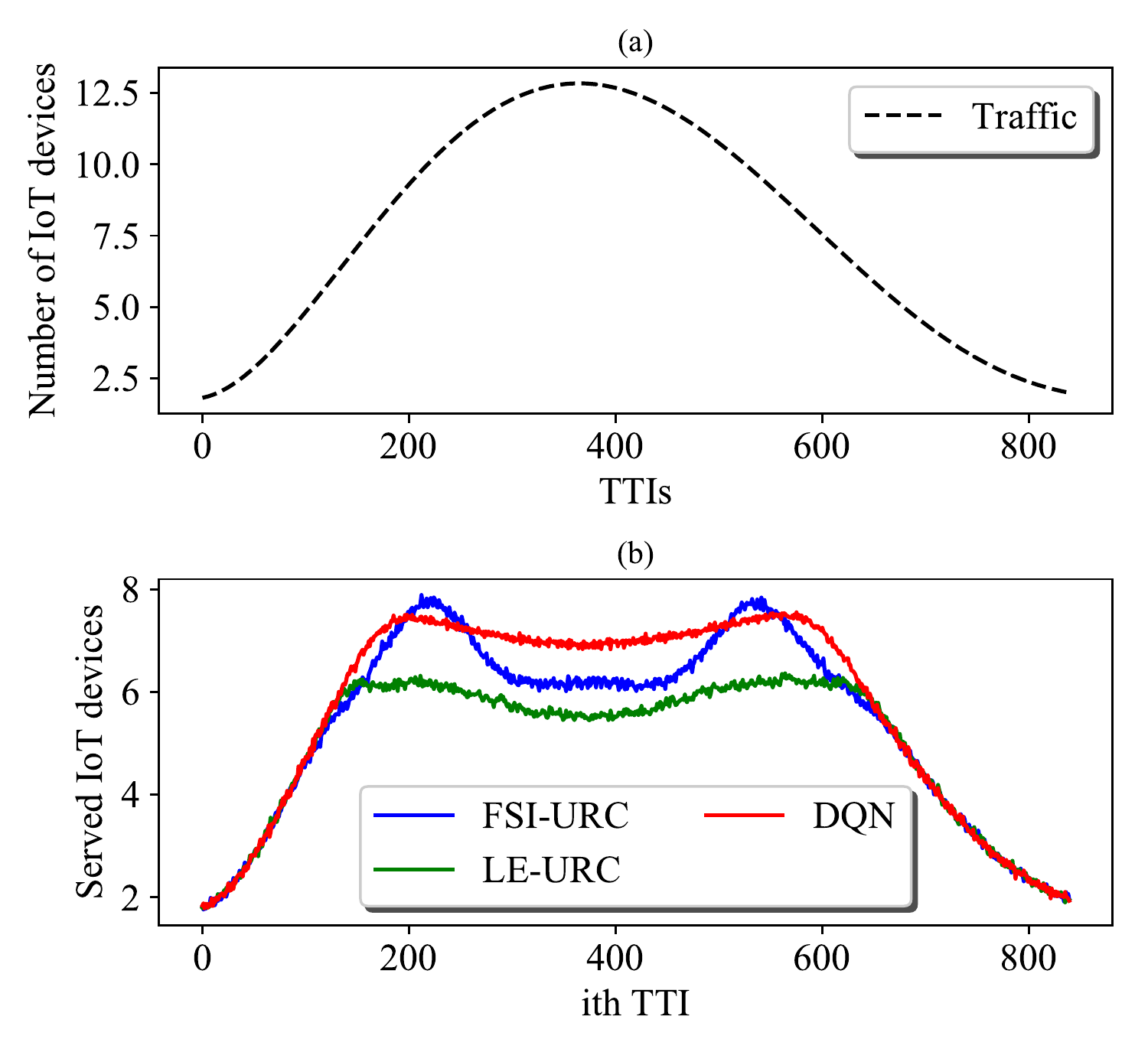}
        \vspace*{-1.2cm}
        \caption{\scriptsize The real-time traffic load and $V_\text{su}$ for FSI-URC, LE-URC, and DQN.}
            \label{fig:3}
    \end{minipage}
    \hspace*{+1cm}
        \begin{minipage}[t]{0.44\textwidth}
    \centering
        \includegraphics[width=1\textwidth]{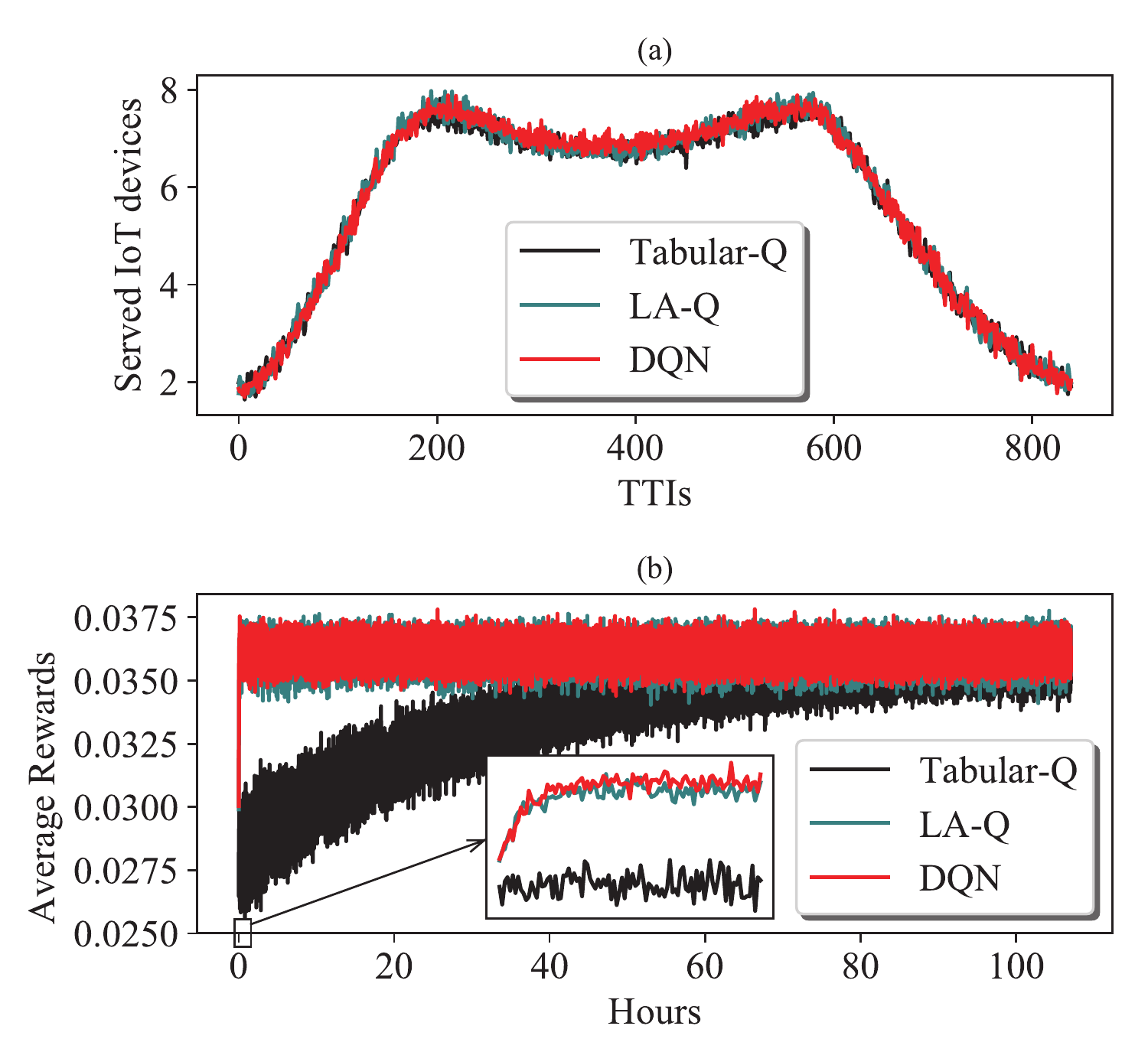}
        \vspace*{-1.2cm}
        \caption{\scriptsize $V_\text{su}$ and the average received reward for Tabular-Q, LA-Q, and DQN.}
                \label{fig:4}
        \end{minipage}
    \end{center}
    \vspace*{-0.5cm}
\end{figure}

Throughout epoch, each device has a periodical traffic profile (i.e., Uniform distritbuion given in Eq. (\ref{q7})), or a bursty traffic profile (i.e., the time limited Beta profile defined in Eq. (\ref{q9}) with parameters $(3,4)$) that has a peak around the 400th TTI. The resulting average number of newly generated packets is shown as dashed line in Fig. \ref{fig:3}(a). Fig. \ref{fig:3}(b) plot the number of successfully served IoT devices $V_\text{su}$ with the proposed FSI-URC, LE-URC, and DQN approaches. In Fig. \ref{fig:3}(b), $V_\text{su}$ first increases gradually with the increasing of traffic shown in Fig. \ref{fig:3}(a), until it reaches the serving capacity of eNB. Then, $V_\text{su}$ decreases slowly due to the increasing collisions and scheduling failures with the increase of traffic. After that, $V_\text{su}$ increases gradually as the collisions and scheduling failures decrease with the decreasing of traffic. Finally, $V_\text{su}$ decreases slowly with the decreasing of traffic.

In Fig. \ref{fig:3}(b), we see that the ideal FSI-URC approach outperforms the LE-URC approach, due to that the FSI-URC approach uses the actual network load to perfectly optimize $V^t_\text{su}$ at one time instance as Eq. (\ref{q6-2}). DQN not only always outperforms LE-URC, but also exceeds the ideal DSI-URC approach in most of TTIs. This is due to that both LE-URC and FSI-URC only optimize $V^t_\text{su}$ at one time instance, whereas DQN optimizes the long-term performance of the number of served IoT devices. The optimization in one time instance (LE-URC and FSI-URC) only takes into account the current trade-off between RACH resource and DATA resource given in Eq. (\ref{1q5}), while the optimization over long-term period (DQN) also accounts for some long-term hidden features, such as the dropping packets due to exceeding them maximum RACH attempts $\gamma_\text{pMax}$ or maximum resource requests $\gamma_\text{RRC}$. The DQN approach can well capture these hidden features to optimize the long-term performance of $V_\text{su}$ as Eq. (\ref{q6-1}).

Fig. \ref{fig:4}(a) compares the number of successfully served IoT devices $V_\text{su}$ under Tabular-Q, LA-Q, and DQN approaches. We observe that all these three approaches achieve similar values of $V_\text{su}$, which indicates that both LA-Q and DQN can well estimate the optimal value function $Q^*(s,a)$ as the converged Tabular-Q in this low-complexity single CE group scenario. Fig. \ref{fig:4}(b) plots the average received reward over each bursty duration ${\mathbb E}\{R\} = \frac{1}{T_\text{bursty}} \sum_{t=0}^{T_\text{bursty}}R^t$ (i.e., one epoch consists of one bursty duration $T_\text{bursty}$) from the beginning of the training versus the required training time. It can be seen that LA-Q and DQN converge to the optimal value function $Q^*(s,a)$ (about 10 minutes) much faster than that of Tabular-Q (about 5 days). The observations in Fig. \ref{fig:4} demonstrate that LA-Q and DQN can be good alternatives for tabular-Q to achieve almost same number of served IoT devices with much less training time.

\begin{figure}[htbp!]
    \begin{center}
    \begin{minipage}[t]{0.44\textwidth}
    \centering
        \includegraphics[width=1\textwidth]{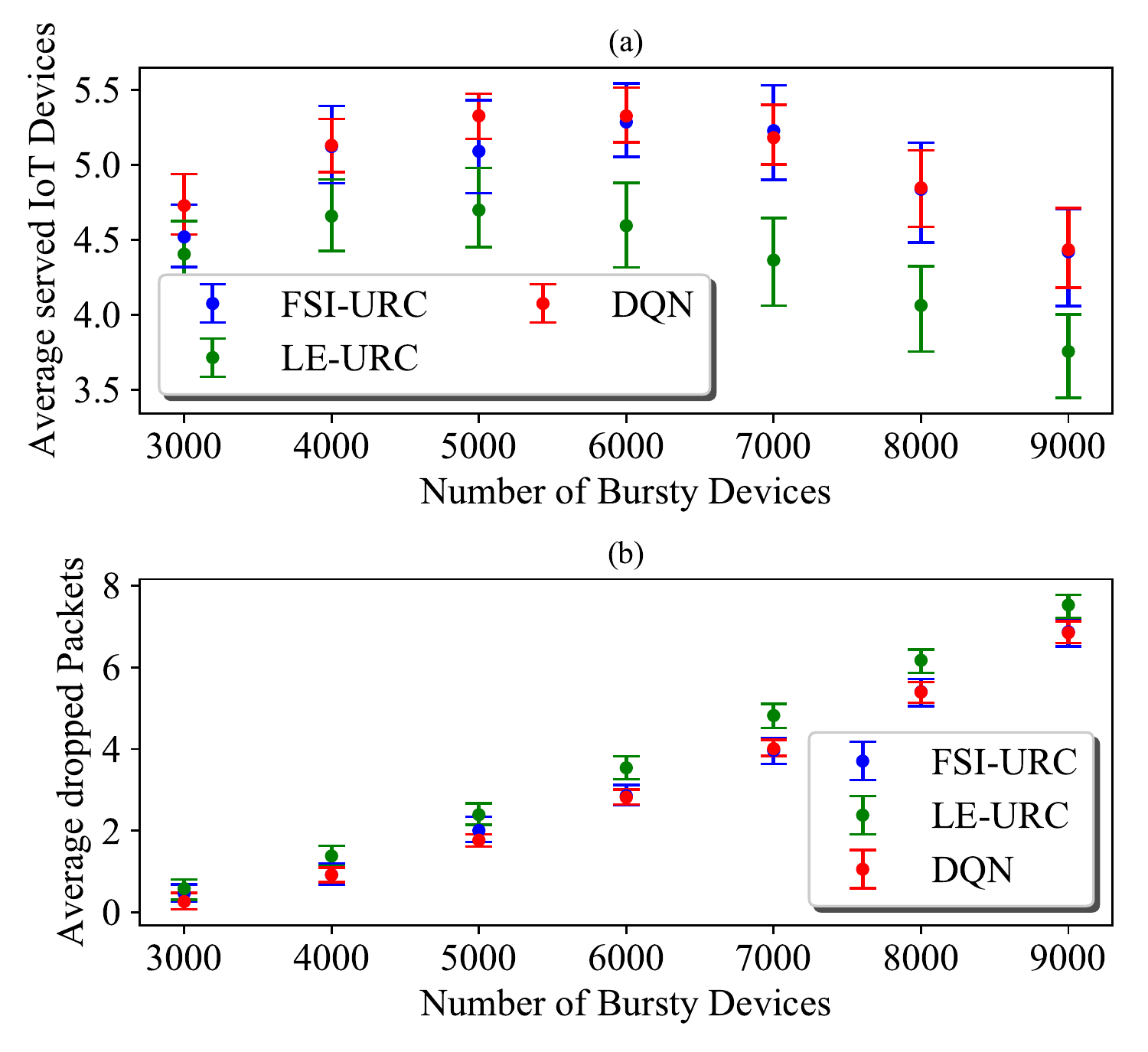}
        \vspace*{-1.2cm}
        \caption{\scriptsize ${\mathbb E}\{V_\text{su}\}$ and ${\mathbb E}\{V_\text{drop}\}$ for FSI-URC, LE-URC, and DQN.}
            \label{fig:5}
    \end{minipage}
    \hspace*{+1cm}
        \begin{minipage}[t]{0.44\textwidth}
    \centering
        \includegraphics[width=1\textwidth]{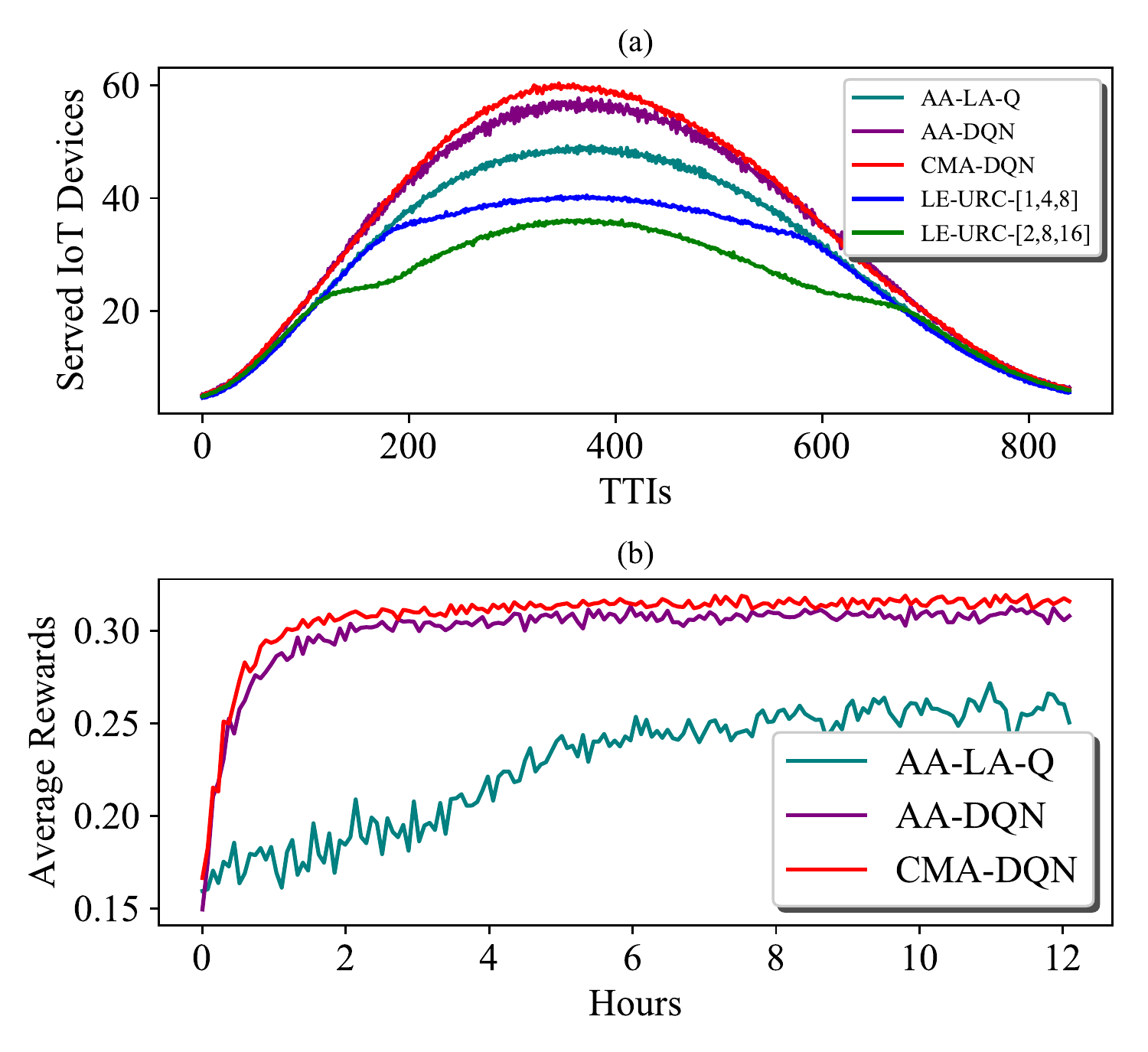}
        \vspace*{-1.2cm}
        \caption{\scriptsize $V_\text{su}$ and the average received reward.}
                \label{fig:6}
        \end{minipage}
    \end{center}
\end{figure}

Fig. \ref{fig:5}(a) and Fig. \ref{fig:5}(b) plot the average number of successfully served IoT devices ${\mathbb E}\{V_\text{su}\}$ and the average number of dropped packets ${\mathbb E}\{V_\text{drop}\}$ (i.e., this system performance can only be summarized in simulation) over a bursty duration $T_\text{bursty}$ versus the number of bursty IoT devices $D_\text{bursty}$. In Fig. \ref{fig:5}(a), we observe that ${\mathbb E}\{V_\text{su}\}$ first increases and then decreases with increasing the number of bursty devices, the decreasing trend starts when eNB can not afford to serve the increasing IoT device number due to the increasing collisions and scheduling failures. These collisions and scheduling failures also result in the increasing number of packet drops with increasing traffics as shown in Fig. \ref{fig:5}(b). In Fig. \ref{fig:5}, we also notice that DQN always outperforms LE-URC (especially for relatively large $D_\text{bursty}$), which indicates the superiority of DQN approach in handling massive bursty IoT devices. Interestingly, DQN provides better performance of the number of served IoT devices and smaller mean errors than the ideal FSI-URC approach in most cases, which thanks to the long-term optimization capability of DQN.

\vspace*{-0.0cm}
\vspace*{-0.0cm}

\subsection{Multi-Parameter Multi-Group Scenario}
Considering eNB is located at the center of a circle area with 12 km radius, we set RSRP thresholds for CE group choosing $\{\gamma_\text{RSRP1}, \gamma_\text{RSRP2}\} = \{0,-5\}$dB, the uplink resource $R_\text{uplink}=15360$ REs (i.e., 320 slots with 48 sub-carriers), and the NPUSCH constrains for LE-URC following $R_{\text{uplink},0}:R_{\text{uplink},1}:R_{\text{uplink},2} = 1:1:1$. To model massive IoT traffic, both the number of periodical IoT devices $D_\text{periodic}$ and the number of bursty IoT devices $D_\text{bursty}$ increase to $30000$. In AA-DQN, we use one Q-network with three hidden layers each of which is consist of 2048 ReLU units. In CMA-DQN, nine DQNs are used to control each of the nine configuration (i.e., $n_{\text{Rach},i}$, $n_{\text{Repe},i}$, $f_{\text{Prea},i}$ for three CE groups), where each DQN has three hidden layers, each with 128 ReLU units. AA-LA-Q and AA-DQN approaches are proposed in Sec. V.A, and CMA-DQN approach is proposed in Sec. V.B.


Fig. \ref{fig:6}(a) compares the number of successfully served IoT devices $V_\text{su}$ during one epoch using AA-LA-Q, AA-DQN, CMA-DQN and LE-URC. The ``LE-URC-[1,4,8]" and ``LE-URC-[2,8,16]" curves represent the LE-URC approach with the repetition values $\{n_{\text{Repe},0},n_{\text{Repe},1},n_{\text{Repe},2}\}$ set to $\{1,4,8\}$ and $\{2,8,16\}$, respectively. We observes that the number of successfully served IoT devices $V_\text{su}$ follows CMA-DQN$\hspace*{+0.1cm}>\hspace*{+0.1cm}$AA-DQN$\hspace*{+0.1cm}>\hspace*{+0.1cm}$AA-LA-Q$\hspace*{+0.1cm}\gg\hspace*{+0.1cm}$LE-URC-[1,4,8]$\hspace*{+0.1cm}\gg\hspace*{+0.1cm}$LE-URC-[2,8,16]. As can be seen, all Q-learning based approaches outperform LE-URC approaches, due to that these Q-learning based approaches can dynamically optimize the number of served IoT devices by accurately configuring each parameter. We also observe that CMA-DQN slightly outperforms the others in the light traffic regions at the beginning and end of the epoch, but it substantially outperforms the others in the period of heavy traffic in the middle of the epoch. This demonstrates the capability of CMA-DQN in better managing the scarce channel resource in the presence of heavy traffic. It is also observed that increasing the repetition value of each CE group with LE-URC improves the received SNR, and thus the RACH success rate in the light traffic region, but it degrades the scheduling success rate due to limited channel resource in the heavy traffic region.

Fig. \ref{fig:6}(b) plots the average received reward over each bursty duration ${\mathbb E}\{R\} = \frac{1}{T_\text{bursty}} \sum_{t=0}^{T_\text{bursty}}R^t$ from the beginning of the training versus the consumed training time. It can be seen that CMA-DQN and AA-DQN outperform AA-LA-Q in terms of less training time. Compared with the results in the single CE group scenario shown in Fig. \ref{fig:4}, DNN is a better value function approximator for the 3 CE groups scenario due to its efficiency and capability in solving high complexity problems. We also observe that CMA-DQN achieves higher ${\mathbb E}\{R\}^* $ than AA-DQN, due to that CMA-DQN can accurately select the exact values from the set of actions $\{{\boldsymbol{\cal N}}_\text{Repe}, {\boldsymbol{\cal N}}_\text{Rach}, {\boldsymbol{\cal F}}_\text{Prea}\}$, whereas AA-DQN can only select ascent/descent actions, which leads to a sub-optimal solution.

\vspace*{-0.0cm}
\begin{figure}[htbp!]
    \begin{center}
        \begin{minipage}[t]{1\textwidth}
    \centering
        \includegraphics[width=1\textwidth]{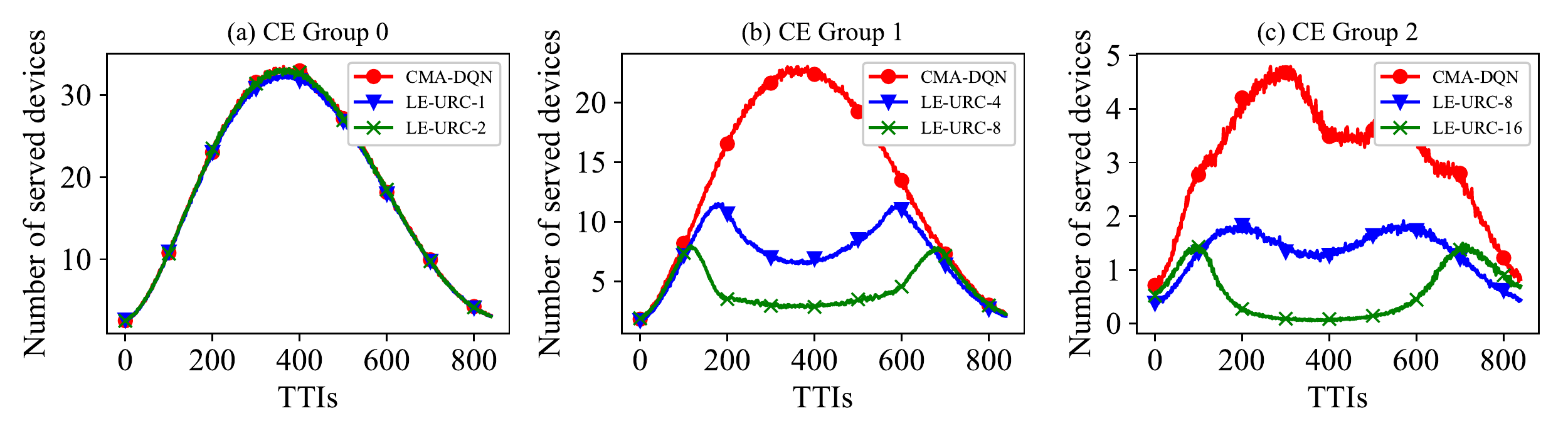}
        \vspace*{-1.2cm}
        \caption{\scriptsize The average number of successfully served IoT devices $V_{\text{succ},i}$ for each CE group $i$.}
                \label{fig:8}
        \end{minipage}
            \begin{minipage}[t]{1\textwidth}
    \centering
        \includegraphics[width=1\textwidth]{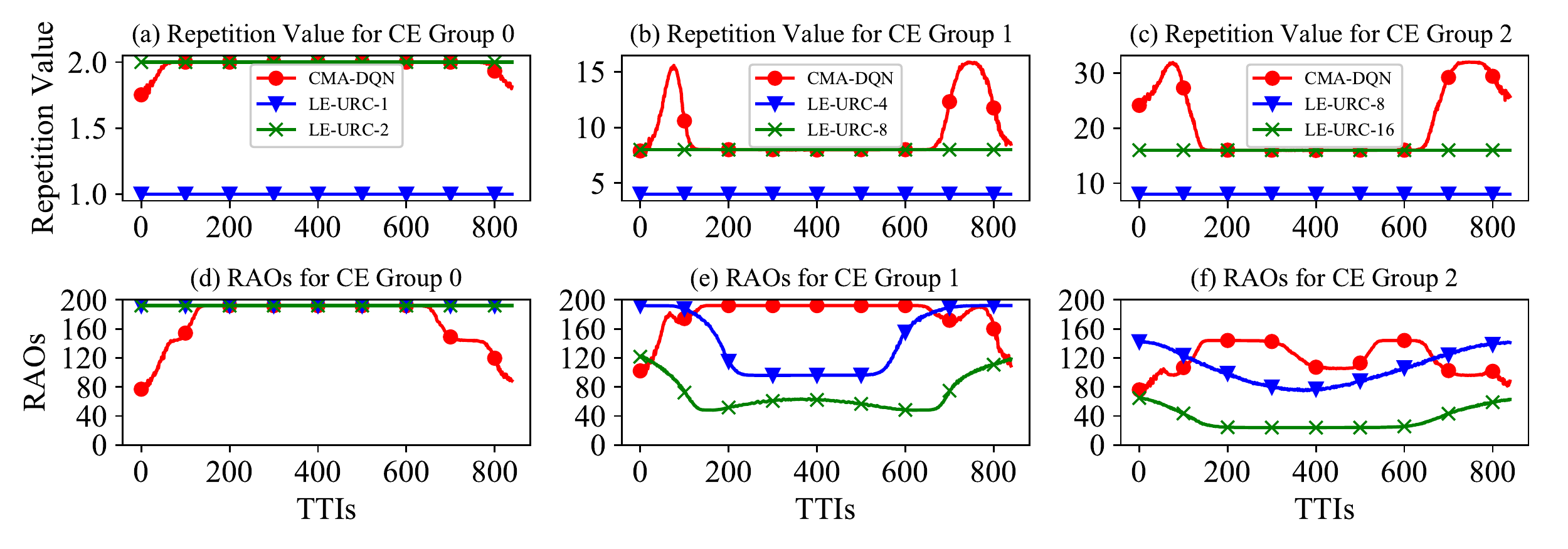}
        \vspace*{-1.2cm}
        \caption{\scriptsize The allocated repetition value $n^t_{\text{Repe},i}$, and RAOs producted by $n^t_{\text{Rach},i} \times f^t_{\text{Prea},i}$.}
            \label{fig:9}
    \end{minipage}
    \end{center}
\end{figure}

To gain more insight into the operation of CMA-DQN, Fig. \ref{fig:8} plots the average number of successfully served IoT devices $V_{\text{succ},i}$ for each CE group $i$, and Fig. \ref{fig:9} plots the average number $n^t_{\text{Repe},i}$ of repetitions and the average number of Random Access Opportunities (RAOs), defined as the product $n^t_{\text{Rach},i} \times f^t_{\text{Prea},i}$, for each CE group $i$ that are selected by CMA-DQN over the testing episodes. As seen in Fig. \ref{fig:8}, CMA-DQN substantially outperforms LE-URC approaches for each CE group $i$, where the reasons for this performance are showcased in Fig. \ref{fig:9}. As seen in Fig. \ref{fig:9}(a)-(c), CMA-DQN increases the number of repetitions in the light traffic region in order to improve the SNR and reduce RACH failures, while decreasing it in the heavy traffic region so as to reduce scheduling failures. Surprisingly, the CMA-DQN increases the repetition value of group 0 $n_{\text{Repe},0}$ at the same time, which is completely opposite to the actions of $n_{\text{Repe},1}$ and $n_{\text{Repe},2}$. This is due to that the CMA-DQN is aware of the key to optimize the overall performance $V_\text{su}$ is to guarantee $V_{\text{succ},0}$, as the IoT devices in the CE group 0 are easier to be served, due to they are located close to the eNB and consume less resource. As illustrated in Fig. \ref{fig:9}(d)-(f), this allows CMA-DQN to increase the number of RAOs in the high traffic regime mitigating the impact of collisions on the throughput. In contrast, for the CE groups 1 and 2, in the heavy traffic region, LE-URC decreases the number of RAOs in order to reduce resource scheduling failures, causing an overall lower throughput as seen in Fig. \ref{fig:8}.

\begin{figure}[htbp!]
\vspace*{-0.0cm}
    \begin{center}
        \begin{minipage}[t]{0.44\textwidth}
    \centering
        \includegraphics[width=1\textwidth]{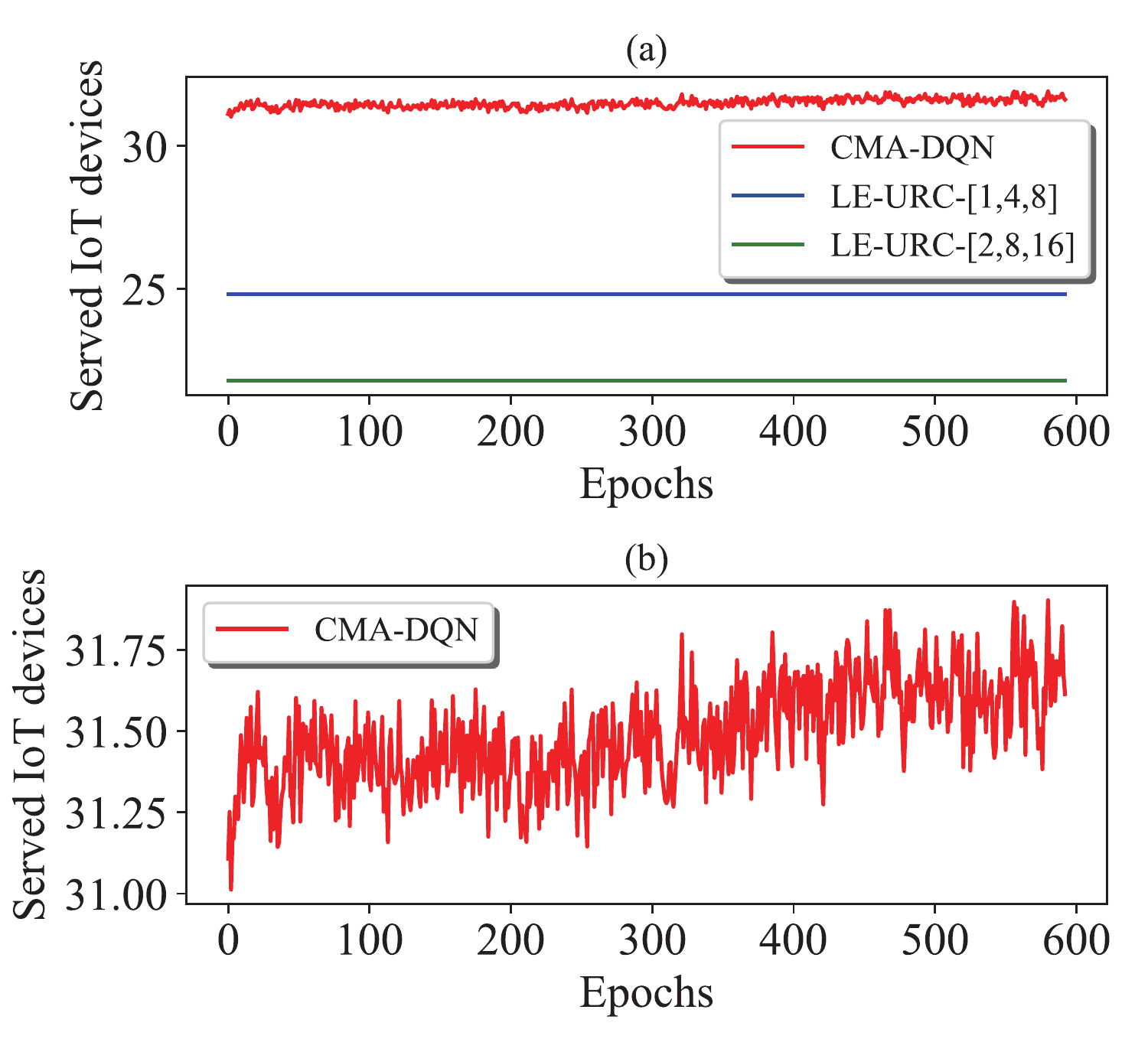}
        \vspace*{-1.2cm}
        \caption{\scriptsize The average number of successfully served IoT devices per TTI over each epoch in online updating}
                \label{fig:10}
        \end{minipage}
    \end{center}
    \vspace*{-0.6cm}
\end{figure}

The realistic network conditions can be different from the simulation environment, due to that the practical traffic and physical channel vary and can be unpredictable. This difference may lead to inaccurate configuration that can degrade the system performance of each approach. Fortunately, the proposed RL-based approaches can self-update after deployment according to the practical observation in NB-IoT networks in an online manner. To model this, we use the trained CMA-DQN agents given in Fig. \ref{fig:9} (i.e., the bursty traffic is modelled by the time limited Beta profile with parameters $(3,4)$), and test them in a slightly modified traffic scenario that the bursty traffic is with Beta$(5,6)$, and we set the constant exploration rate $\epsilon=0.001$. Fig. \ref{fig:10} plots the average number of successfully served IoT devices ${\mathbb E}\{V_\text{su}\}$ per TTI over each episode versus epochs. Our result shows that, as expected, ${\mathbb E}\{V_\text{su}\}$ follows CMA-DQN$>$LE-URC-[1,4,8]$>$LE-URC-[2,8,16] at any epoch. More importantly, the performance of CMA-DQN gradually improves along epochs, which sheds light on the online self-updating capability of the proposed RL-based approaches.

\vspace*{-0.3cm}

\section{CONCLUSION}
\vspace*{-0.1cm}
In this paper, we developed Q-learning based uplink resource configuration approaches to optimize the number of served IoT devices in real-time in NB-IoT networks. We first developed tabular-Q, LA-Q, and DQN based approaches for the single-parameter single-group scenario, which are shown to outperform the conventional LE-URC and FSI-URC approaches in terms of the number of served IoT devices. Our results demonstrated that LA-Q and DQN can be good alternatives for tabular-Q to achieve almost the same system performance with much less training time. To support traffic with different coverage requirements, we then studied the multi-parameter multi-group scenario as defined in NB-IoT standard, which introduced the high-dimensional configurations problem. To solve it, we advanced the proposed LA-Q and DQN using the Actions Aggregation technique (AA-LA-Q and AA-DQN), which guarantees the convergent capability of Q-learning by sacrificing the accuracy in resource configuration. We further developed CMA-DQN by dividing high-dimensional configurations into multiple parallel sub-tasks, which achieved the best performance in terms of the number of successfully served IoT devices $V_\text{su}$ with the least training time.

\appendices
\numberwithin{equation}{section}

\ifCLASSOPTIONcaptionsoff
  \newpage
\fi

\bibliographystyle{IEEEtran}
\bibliography{IEEEabrv,RA_bib}

\end{document}